\documentclass{aa}  

\usepackage{graphicx}
\usepackage{txfonts}
\usepackage{colortbl}
\usepackage{hyperref}
\hypersetup{colorlinks=true,linkcolor=[rgb]{1.,0.2,0.2},citecolor=[rgb]{0.1,0.1,1.},filecolor=[rgb]{0.7,0.2,0.2},urlcolor=[rgb]{0.7,0.2,0.2}}

\usepackage{pifont}
\def \Jbhv{$\vec{J}_{\rm BH}\,$}
\def \Jdv{$\vec{J}_{\rm d}\,$}

\newcommand{\msun}{\,\mathrm{M_\odot}}

\newcommand{\mstar}{\,\mathit{M_*}}
\newcommand{\mhalo}{\,\mathit{M}_{\rm halo}}
\newcommand{\lgal}{\texttt{L-Galaxies}}
\newcommand{\lgbh}{\texttt{L-Galaxies}\textit{BH}}
\newcommand{\supereddington}{\texttt{SuperEdd}}
\newcommand{\eddington}{\texttt{EddLim}}
\newcommand{\heavyMax}{\texttt{EddLim-HeavyMax}}
\newcommand{\heavyMin}{\texttt{EddLim-HeavyMin}}

\newcommand{\graft}{\textit{grafting}}

\newcommand{\MS}{MS}
\newcommand{\MSII}{MSII}
\newcommand{\MSG}{MS+\texttt{Grafting}}

\definecolor{mypurple}{rgb}{0.6, 0.0, 0.3}

\usepackage{color}
\definecolor{myorange}{rgb}{0.8, 0.3, 0.0}

\definecolor{mygreen}{rgb}{0.1, 0.55, 0.175}

\definecolor{myred}{rgb}{0.1, 0.55, 0.175}

\begin{document} 

  \title{Constraints on the early growth of massive black holes from PTA and JWST with \lgbh{}}

  % \subtitle{Galaxy formation models make PTA and JWST speak to each other}

   \author{ Silvia Bonoli$^{*1,2}$, David Izquierdo-Villalba$^{3,4}$, Daniele Spinoso$^{5}$, \\ Monica Colpi$^{3,4}$, Alberto Sesana$^{3,4,6}$, Markos Polkas$^{1,7}$, Volker Springel$^{8}$
          }

 \institute{$^{1}$ Donostia International Physics Centre (DIPC), Paseo Manuel de Lardizabal 4, 20018 Donostia-San Sebastian, Spain \\ 
     $^{2}$ IKERBASQUE, Basque Foundation for Science, E-48013, Bilbao, Spain \\ 
      $^{3}$ Dipartimento di Fisica ``G. Occhialini'', Universit\`{a} degli Studi di Milano-Bicocca, Piazza della Scienza 3, I-20126 Milano, Italy\\
     $^{4}$ INFN, Sezione di Milano-Bicocca, Piazza della Scienza 3, 20126 Milano, Italy\\  
     $^{5}$ Como Lake Center for AstroPhysics,  University of Insubria, 22100, Como, Italy\\ 
     $^{6}$ INAF - Osservatorio Astronomico di Cagliari, via della Scienza 5, 09047 Selargius (CA), Italy\\ 
     $^{7}$ University of the Basque Country UPV/EHU, Department of Theoretical Physics, Bilbao, E-48080, Spain\\ 
     $^{8}$ Max-Planck-Institut für Astrophysik, Karl-Schwarzschild-Str. 1, 85748 Garching, Germany \\ \\
   \email{silvia.bonoli@dipc.org}}

   \date{}

% \abstract{}{}{}{}{} 
% 5 {} token are mandatory
 
  \abstract{
   Recent Pulsar Timing Arrays (PTAs) results provided strong evidence for a stochastic gravitational wave background (sGWB), consistent with a population of merging massive black holes (MBHs) at $z<1$. Meanwhile, JWST observations at $z>5$ suggest a higher number density of accreting MBHs than previously estimated. Together with constraints from local MBHs and high-$z$ quasars, these findings offer a unique opportunity to test MBH seeding and early growth models. We explore this using  {\tt L-Galaxies}\textit{BH}, a new  extension of the galaxy formation model {\tt L-Galaxies}, developed to explicitly model all stages of MBH evolution, including seeding, accretion, and binary dynamics. To take advantage of both the high resolution of the {\tt MillenniumII} and the large volume of the {\tt Millennium} simulations, we run {\tt L-Galaxies}\textit{BH} on the former and use its outputs as initial conditions for the latter, via our \textit{grafting} method. We find that reproducing the number density of high-$z$ active MBHs observed by JWST requires either a heavy seed formation rate significantly higher than that predicted by current predictions ($\gtrsim 0.01 \rm{Mpc}^{-3}$ at $z \sim 10$), or widespread formation of light seeds undergoing multiple phases of super-Eddington accretion. Furthermore, matching the amplitude of the PTA sGWB signal requires nearly all galaxies with stellar masses $M_{*}> 10^9 M_\odot$ to host central MBHs by $z\sim0$. Given the extreme heavy seed densities required to satisfy both PTA and JWST constraints, our results favor a scenario in which MBHs originate from light seeds that grow rapidly and efficiently in the early universe. This work demonstrates the power of combining multi-messenger data with physical models to probe the origins and evolution of MBHs across cosmic time.

  }

   \keywords{Methods: numerical --- quasars: supermassive black holes -- Gravitational waves -- Galaxies: interactions }
   \titlerunning{Early growth of MBHs with JWST, PTA and \lgbh{}}
    \authorrunning{Bonoli et al.}
   \maketitle
%
%-------------------------------------------------------------------

\section{Introduction}

With the detection of the first gravitational waves (GWs) from  the coalescence of two stellar-mass binary black holes \citep{Abbott2016}, the Advanced Laser Interferometer Gravitational-Wave Observatory (LIGO) has opened a new era into the study of extreme compact objects, complementing electromagnetic  observations.  
While ground based interferometers have sufficiently long baselines to detect GWs from stellar-size black holes, the detection of the inspiral and merger of more massive ($> 10^3 \msun$) black holes requires much longer baselines.  
The Laser Interferometer Space Antenna (LISA), with planned launch in the mid-2030s \citep{LISA2024}, will be able to probe the GWs emitted by coalescing black holes in the $10^3\,{-}\,10^7 \msun$ mass range. Reaching up to very high redshifts, LISA will potentially provide unique constraints on the early growth and evolution of massive black holes (MBHs), that will be complemented by observations in the electromagnetic realm  \citep[e.g., ][]{AmaroSeoane2023}.  

Moving to the most massive black holes, variations in the arrival times of the signal from millisecond pulsars can be due to the stretching of space-time caused by the passing of GWs.   \cite{Hellings1983}  predicted that a stochastic gravitational-wave background (sGWB) would lead to a correlation in pulsar timing residuals as a function of the pulsars angular separation (Hellings and Downs curve). The Pulsar Timing Array (PTA) collaborations EPTA-InPTA, NanoGrav, PPTA and CPTA collaborations have recently reported a highly significant signal of a sGWB \citep{EPTA2023A&A_primary,Agazie2023ApJ_primary,Reardon2023,Xu2023}. While the shape of the strain signal would be consistent with the one expected from the merger of very massive black holes \citep[i.e., black holes with mass $\rm {\gtrsim}\,10^8\, \rm \msun$, ][]{EPTA2024_BHs, Agazie2023ApJSMBHB}, its amplitude is larger than the one predicted by models of the evolution of massive binary black holes \citep[e.g.,][]{Sesana2013}. 
By leveraging its dependence on the masses of merging MBHs, the amplitude of the sGWB signal %detected with pulsar timing
can indeed be used to constrain models for the growth of MBHs. \citet{IzquierdoVillalba2021}, for example, used early PTA limits to conclude that the number density of black holes with mass $\rm 
 M_{BH} \,{\gtrsim}\, 10^{8} \, \msun$ must be larger than current estimates to reach a large enough sGWB amplitude \citep[see also ][]{SatoPolito2023, LiepoldAndMa2024}. 

On the electromagnetic side, the James Webb Space Telescope  (JWST) is now providing revolutionary observations: peering into the distant Universe,  the discovery of an abundant population of sources consistent with low-luminosity AGN is offering new insights into the early assembly of MBHs. Surveys such as FRESCO,  UNCOVER,  JADES, and RUBIES reveal a population of accreting MBHs at $z \,{\gtrsim}\, 4 $ with masses   $\rm 
 10^{6} \, {\lesssim} \, M_{BH} \, {\lesssim}\, 10^{8} \, \msun$ and number densities around $10^{-4}\,{-}\,10^{-5} \, \rm Mpc^{-3}$ \citep[][]{Kokorev2023, Matthee2024, Greene2024, Maiolino2024, DeGraaf2025} and even as high as $10^{-3}\,{-}\,10^{-2} \, \rm Mpc^{-3}$ for the least luminous objects \citep[][]{Geris2025}. 
 Some of these samples include the so-called Little Red Dots (LRDs), a population of compact galaxies characterized by a puzzling  ``v-shaped'' UV-to-optical rest-frame continuum as well as (in most cases) broad spectral lines, suggesting the presence of an accreting MBH \citep[][]{Matthee2024, Greene2024}. Current explanations for LRDs range from direct collapse black holes in the process of formation \citep[][]{Pacucci2025, Zwick2025, Jeon2025, Cenci2025}, tidal disruption events around an infant black hole \citep{Bellovary2025, Perger2025}, accretion around primordial black holes \citep[e.g.,][]{Ziparo2025} to black hole seed formation driven by self-interacting dark matter  \citep{Roberts2025}. The spectral shape of LRDs and their extreme faintness in X-ray could possibly be explained if these black holes are accreting above the Eddington limit   \citep{Liu2025, Inayoshi2024SE, Madau2024, InayoshiMaiolino2025, Madau2025}. While still uncertain, current estimates of the masses of these black holes and the ones of their host galaxies put them either above the local scaling relation \citep[e.g.,][]{Kokorev2023, Harikane2023, Juodzbalis2025} or closer to the local relations when considering spectral changes if emitting close or above the Eddington limit \citep{Lupi2024} or if non-standard bolometric corrections are used \citep{Greene2025}.

A few theoretical works have already used these JWST results to study the assembly of MBHs. For example, \cite{Li_Inayoshi2024} model and follow MBH  seeding and evolution at high-$z$ ($z\,{>}\,4$) with Press-Schechter \citep[][]{PressANDSchechter1974}  merger trees to interpret the JWST population of active black holes, finding that their parameters that regulate MBH growth are highly redshift-dependent, with rapid growth at $z\,{>}\,6$ decelerating at lower-redshift. \cite{trinca2023} and \cite{trinca2024} also explore the growth of MBHs via the semi-analytical model CAT run on Press-Schechter merger trees and conclude that super-Eddington accretion is likely the reason for the fast build-up of these objects. While these studies focus on electromagnetic constraints from high-$z$ observations, the multi-messenger perspective remains limited. Indeed, multi-messenger
studies on single and binary MBHs are still not possible as
PTA collaborations have not yet discovered the GW emission
from individual sources (i.e continuous GWs). 
%While these studies focus on electromagnetic constraints from high-$z$ observations, the multi-messenger perspective remains limited. 
%This is primarily due to the lack of detections of GW emission from individual sources (i.e., continuous GWs) by PTA collaborations.

However, the recent PTA results about the sGWB can  be combined with JWST observations, offering very interesting \textit{statistical} multi-messenger constraints to test models for the formation and evolution of MBHs in a cosmological context. Along these lines, \citet{Toubiana2024} have used these new datasets to explore the physics of MBHs within the \texttt{POMPOCO} parametric model, which follows the evolution of MBHs within Press-Schechter merger trees using 12 free parameters. The authors found that heavy seeds, rapid black hole mergers and super-Eddington phases are needed to reconcile the observations. A similar analysis was also carried out by \citet{Ellis2024}.

Combining electromagnetic and GW constraints and interpreting them jointly to solve the puzzle of the seeding and cosmological evolution of MBHs is a challenging endeavor.

It involves modelling the birth of the first black hole seeds within the earliest-forming halos, understanding the physical processes that drive their co-evolution with host galaxies, and tracking their dynamical evolution when they pair with other black holes and eventually release GWs \citep[e.g., ][]{Volonteri2021NAT}. Addressing this challenge demands embedding MBH evolution within comprehensive galaxy formation models capable of self-consistently following all key stages (seeding, growth, and dynamics) while employing a broad range of electromagnetic and GW observations to constrain the possible evolutionary pathways, given the high degree of degeneracy inherent in this multiscale framework.
Furthermore, this comprehensive modeling must be used to simulate wide, cosmological volumes in order to obtain statistically significant populations of galaxies and MBHs. This is a key aspect, which often limits the feasibility of such analysis in terms of computational costs, as well as the possibility to reliably compare theoretical results with observational data.

In this work, we use our new \lgbh{} model to interpret the evolution of MBHs and MBHBs by utilising the latest results from PTA and JWST. \lgbh{} is a new branch of the semi-analytical model (SAM) for galaxy evolution \lgal{}, built upon the recent works of \citet{IzquierdoVillalba2020, IzquierdoVillalba2021, Spinoso2022, Polkas2024}. It has been developed precisely to study with a high level of detail the physical processes that drive the formation, growth and mergers of MBHs within their host galaxies and dark matter halos. Following the structure of \lgal{}, \lgbh{} is also designed to run on merger trees extracted from cosmological N-body simulations. This is one of the main strengths of \lgal{}, as N-body simulations provide a more accurate description of the dark matter halo population over time, including merger timescales, number densities, and environmental effects, compared to Press-Schechter merger trees. In this work, we make use of both the \texttt{Millennium} \citep[\MS{},][]{Springel2005} and \texttt{Millennium-II} \citep[\MSII{},][]{Boylan-Kolchin2009} simulations. Specifically, we extend the merger trees of the \MS{} using those from the \MSII{}, thereby exploiting the larger box and statistics of the former while taking advantage of the higher resolution of the latter. This approach is crucial for accurately capturing the early phases of black hole formation and growth, deriving reliable statistics of MBH merger events, and tracking the assembly of the most massive black holes and their host galaxies. Within this framework, we 
%study 
are able to follow \textit{self-consistently}
the evolution of MBHs from the formation of their seeds to the growth of billion solar-mass black holes, jointly employing the latest PTA and JWST results to constrain the physical processes driving their evolution. Our focus lies in exploring how current and upcoming \textit{statistical} multi-messenger datasets (that is, GW data combined with traditional electromagnetic observations of large MBH samples) can be employed to constrain the properties of the MBH population across a wide mass range and provide new insights into their formation and evolutionary pathways.\\

The paper is organized as follows. In Section~\ref{sec:lgbh} we start by reviewing the main properties of the DM merger trees we use and describe the \textit{grafting} procedure that we developed to combine the merger trees of multiple simulations. We then summarize the basic aspects of \lgal{} and the details of the physics of MBHs included in our new \lgbh{}. In Section~\ref{sec:ALL_model_variations} we present the four model variations that we test in this work.  In Section~\ref{sec:model_constraints}, we use JWST and PTA data to constrain the degeneracies between the model predictions. Details on the preferred models are presented in Section~\ref{sec:prefModels}. Finally, in Section~\ref{sec:summary} we summarize and discuss the main results of the paper. Throughout the paper we adopt a Lambda Cold Dark Matter $(\Lambda$CDM) cosmology with parameters $\Omega_{\rm m} \,{=}\,0.315$, $\Omega_{\rm \Lambda}\,{=}\,0.685$, $\Omega_{\rm b}\,{=}\,0.045$, $\sigma_{8}\,{=}\,0.9$ and $h \, {=} \, \rm H_0/100\,{=}\,67.3/100\, \rm km\,s^{-1}\,Mpc^{-1}$ \citep[from][]{PlanckCollaboration2014}.

%--------------------------------------------------------------------

\section{\lgbh: from Dark Matter merger trees to MBHs populations} \label{sec:lgbh}

In this section, we provide a detailed overview of our model. We start by discussing the \MS{} and \MSII{} N-body simulations and the DM merger trees that constitute the backbone of the model. Next, we introduce the \textit{grafting procedure}, which allows us to extend the merger history of the dark-matter (DM) trees of the \MS{} by incorporating data from the higher resolution \MSII. We then present a general overview of the \lgal{} semi-analytical model, along with a more in-depth explanation of \lgbh{}, a specific branch of the former model designed to capture all the essential physics involved in the formation, growth, and mergers of MBHs. The physical processes modeled in \lgbh{} have been previously described separately in \citet{Bonoli2009, Bonoli2014, IzquierdoVillalba2020, IzquierdoVillalba2021, Spinoso2022} and \cite{Polkas2024}.

\subsection{Merger trees and dark matter simulations}
DM halo merger trees are the backbone of semi-analytical models of galaxy formation. As for \lgal{}, the \lgbh{} branch has been developed to run on merger trees from N-body cosmological simulations, in particular the \textit{Millennium} suite. In this work, we use the merger trees from the \MS{} and \MSII{} simulations. The \MS{} follows the cosmological evolution of $2160^3$ DM particles with mass ($m_p$) of $8.6\,{\times}\, 10^8\, \msun/h$ inside a periodic box of $500 \, {\rm Mpc}/ h$ on a side, from $z\,{=}\,127$ to the present. The \MSII{} uses the same number of particles than \MS{} but in a box 125 times smaller in volume ($\mathrm{100\,Mpc}/h$) and a mass resolution 125 times higher ($m_p\,{=}\,6.885\,{\times}\,10^6\,\msun/h$).  The simulation outputs of \MS{} and \MSII{} were stored at 63 and 67 epochs, respectively, where the halos were identified by applying friend-of-friend and \texttt{SUBFIND} algorithms and then arranged in \textit{merger trees} with the \texttt{L-HALOTREE} code \citep{Springel2001}. The two simulations were originally run with the WMAP1 \& 2dFGRS concordance cosmology, but recently they were re-scaled with the procedure presented in \cite{AnguloandWhite2010} to match the cosmological parameters provided by Planck first-year data \citep{PlanckCollaboration2014}.

Considering only structures of at least 20 bound particles, the minimum halo masses reached by the \MS{} and \MSII{} simulations are $1.85\,{\times}\, 10^{10}\, \msun/h$ and  $1.53\,{\times}\, 10^8\, \msun/h$, respectively. Taking into account these numbers, the \MS{} allows to study ${\sim}\, \mathit{M}*$ galaxies with large statistical power, and also track the evolution of galaxies in some large clusters (above ${\gtrsim}\, 10^{15}\, \msun/h$). Instead, the \MSII{} offers the mass resolution required to track the history of galaxies down to the dwarf regime (${\lesssim}\,10^{9}\, \msun/h$). In terms of black hole masses, the \MS{} allows for a statistical study of ${\gtrsim}\, 10^6\msun$  MBHs evolution across cosmic time \citep[see e.g., ][]{Marulli2008, Bonoli2009,IzquierdoVillalba2020}, although its volume is only marginally large enough to trace the history of the brightest quasars. Conversely, the \MSII{} can be used to study the evolution of intermediate-mass  ($\rm 10^4\,{-}\,10^6\, \rm \msun$) black holes and the formation sites of MBH seeds \citep[see][]{Bonoli2014, Spinoso2022}.\\

While for most studies the usage of one simulation with the appropriate resolution is generally good enough, that is not the case when one needs to span a wide dynamical range. For instance, investigating the growth of the largest MBHs from their seed masses requires tracking DM halos across a large mass spectrum: from $ 10^6\,{-}\, 10^8 \, \mathrm {\msun}/h$, where first MBHs are thought to have formed, to $ 10^{12}\,{-}\,10^{13} \mathrm {\msun}/h$ which hosts the bright quasars. Moreover, to make accurate predictions of MBH merger rates relevant for current and future GW experiments, it is crucial to comprehensively follow all mergers undergone by the DM halos hosting these black holes. To overcome this situation, in this work we combine the \MS{} and \MSII{} merger trees via a technique that we call \graft. This allows us to reach the higher resolution of the \MSII{} within the larger volume (and statistics) of the \MS{}. In the future, we plan to apply the same procedure to other simulations to reach an even larger dynamical range. In the next section, we introduce the \graft{} technique.\\ 

\begin{figure*}
    \centering
    \includegraphics[width=0.8\columnwidth]{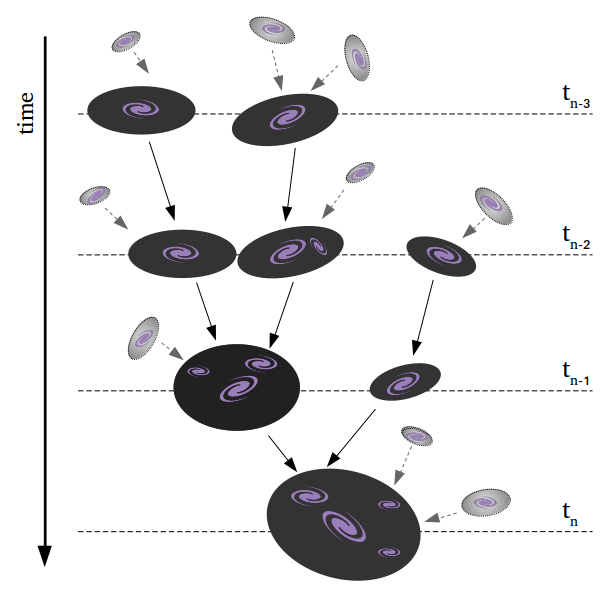}
    \includegraphics[width=1.0\columnwidth]{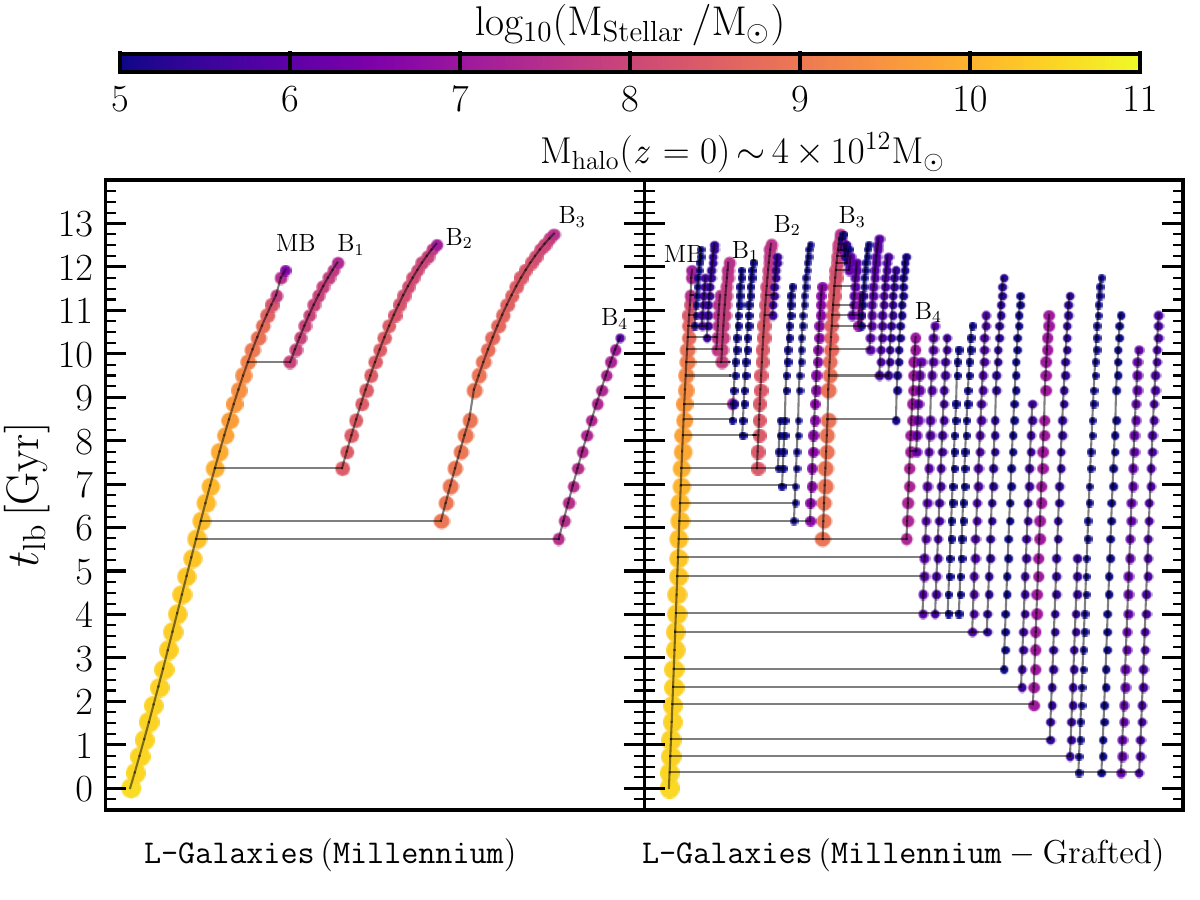}
    \caption{\textbf{Left panel}: Schematic representation of the extension of the merger trees done with the \graft{} procedure. Dark gray ovals indicate the merger tree extracted from the \MS{}, while the light gray, smaller ovals show the progenitors unresolved by the \MS{}, and added as new ``leaves'' by using halos resolved in the \MSII{}. \textbf{Right panel}: An example of a  \textit{galaxy} merger tree of a \MS{} galaxy (left side) and the same \textit {galaxy} merger tree with the added ``leaves'' from the \MSII{} (right side). The halo hosting this galaxy has a mass of ${\mathit{M}_{halo}}(z=0)\, \sim 4 \times 10^{12}\, {\rm M}_{\odot}$. Symbols are color coded  based on the galaxy stellar mass. We underline that this panel shows the \textit{galaxy} merger tree, instead of the halo one. The new halo ``leaves'' are inserted in the merger tree and grafted at the times corresponding to the upper tip of the new galaxy-tracks. After the grafting moment, some galaxies may survive for extended periods of time because of the long dynamical friction timescales before they merge with the central galaxy.} 
    \label{fig:MTE}
\end{figure*}

\subsection{Extending N-body merger trees: The \textit{grafting} technique}
\label{sec:extending_nbody_merger_trees}

One of the main limitations when working with galaxy formation models is the minimum halo mass ($ M_{h}^{Res}$) of the underlying DM merger trees. For models based on N-body simulations, such as \lgbh{}, this limit effectively depends on the particle mass of the simulation, as a halo is considered resolved when a certain number $\mathit{n_{h}^{Res}}$ of particles form a bound system (in our case, $\mathit{n}_{h}^{Res}\,{=}\,20$, so that $\mathit{M}_{h}^{Res}\,{=}\,20\,m_p$).  Thus, (i) the mass of newly resolved objects is always larger than or equal to $M_{h}^{Res}$ and (ii) mergers/interactions where one of the two systems is below  $\mathit{M}_{h}^{Res}$ can not be accounted for. 

The first point imposes a strong limit on the possibility of following galaxy evolution from its early phases, as newly-resolved DM halos could already host an evolved galaxy when the simulation identifies them for the first time\footnote{This is of course not the case if $\mathit{M}_{h}^{Res}$ is small enough that the first collapsing halos where gas can cool and form stars are properly simulated. }. 
The second influences the overall simulated evolution of galaxies, since it prevents to access to information about minor mergers or smooth accretions of small galaxy companions. 

All these shortcomings are unavoidable given the current difficulties in simulating large cosmological volumes at high resolution. In this work, we follow the methodology presented in \cite{Angulo2014}, which aims at mitigating all the flaws derived from the limited resolution of the merger trees extracted from large N-body simulations. In particular, we call this methodology \textit{grafting} and we consider it as a two-step process: the first step aims at rebuilding the accurate number of mergers which would be otherwise lost due to mass resolution, while the second steps aims at reconstructing the unresolved evolution of structures so that newly initialized galaxies would begin to be modeled as evolved systems rather than pristine baryonic reservoirs. In the following, we describe the procedure used in each step. With ``target'' simulation we refer to the N-body simulation with larger volume and lower resolution. With ``input'' simulation we refer instead to the smaller-volume, higher resolution one.

\begin{description}

\item [{\bf Merger tree extension}] To extend the branches of the target simulation below its resolution limit it is necessary to use the information extracted from merger trees of an input higher-resolution simulation\footnote{Note that the time resolution of the high-resolution input simulation must be at least as good as the one featured by the target simulation.}. For a given halo mass bin, $\delta \log_{10}(M_h)\,{=}\,\log_{10}(M_h) \,{\pm}\,\Delta \log_{10}(M_h)$,  and redshift bin, $\delta z \,{=}\, z\,{\pm}\,\Delta z$ of the input simulation, we first determine the average number of halo mergers $\mu$ involving satellites with masses $\delta \log_{10} (m) \,{=}\,\log_{10}(m)\,{\pm}\, \Delta \log_{10}(m)$: 
\begin{equation}
    \bar{\mu}\,{=}\,\mu(\delta \log_{10}(m)\,{|}\,\delta \log_{10}(M_h), \delta z)\ ,
    \label{eq:mu_mean}
\end{equation}

where $\log_{10}(m)$ spans the dynamical range within which the high-resolution simulation can actually provide additional information to the target low-resolution simulation. Thus, $\log_{10}(m)$ varies between the minimum halo mass of the input simulation and the minimum halo mass of the target simulation. Indeed, above its minimum halo mass, the target simulation is able to self-consistently resolve halo mergers. Similarly, the quantity $\log_{10}(M_h)$ spans the mass interval where the dynamical ranges of the high-resolution and target simulations overlap: $\log_{10}(M_h)$ varies between the minimum halo mass of the target simulation and the maximum halo mass of the input high-resolution one. 

The information of $\bar{\mu}$ is then used to extend the merger trees of the target simulation:  for each halo with mass $M'_h \,{\in}\, \delta \log_{10}(M_h) $ and redshift $z'\,{\in}\,\delta z$, an extra number of ``leaves'', i.e., satellite halos of masses $\delta \log_{10} (m)$, are added to the merger trees of the target simulation. The mass distribution of the new satellite halos follows a Poisson distribution with mean $\bar{\mu}$ (see Eq.\ref{eq:mu_mean}). Following \cite{Angulo2014}, the distance of these new satellites with respect to their central halo follows a Lorentz distribution with a mean equal to the virial radius of the central halo.\\

In the left panel of Fig. \ref{fig:MTE} we show a schematic representation of this process when applied to the \MS{} (target) and \MSII{} (input) simulations: the progenitors of selected halos in the \MS{} are not only structures resolved in the \MS{} simulation (dark ovals), but also those added from the \MSII{} (light gray ovals). In the right panel of the same figure, we show both a galaxy merger tree derived running \lgbh{} on the \MS{} alone (left side) and the same merger tree with the addition of new ``leaves'' derived from the \lgbh{} run on the \MSII{}. Clearly, the merger history of the galaxy when including unresolved branches is much richer and more complex than the one followed by the \MS{} alone.\\

 \item [{\bf Halo initialization}] Once the merger trees have been extended, it is necessary to initialize the galaxies inside newly resolved halos. Note that the latter include both the original halos of the target simulation and the additional leaves derived from the input simulation. By default, \lgal{} and \lgbh{} populate newly resolved DM halos with a baryonic content given by the cosmic baryon fraction. This initial baryonic component is included in the form of hot gas which eventually cools and fuels star-formation as time evolves. In principle, any evolution of the baryonic component happening before the redshift in which halos are newly-resolved is neglected. 

 In order to model this un-resolved evolution, on top of adding new DM ``leaves'' to the merger trees of the target simulation, our \graft{} procedure also populates them with an evolved baryonic content, including MBHs. This baryonic counterpart of DM halos is directly taken from a random halo of the same redshift and mass range (within a $\Delta \log_{10}(m_h)$ tolerance) from the outputs of \lgbh{} run on the merger trees of the input high reolsution simulation. This approach has been previously used by \citet{Angulo2014} to populate new halos in the \texttt{Millennium-XXL} simulation as well as by \citet{Bonoli2014} and \citet{IzquierdoVillalba2020} to populate new halos in the \MS{} (in the latter cases, by using the predictions of \MSII{} in a similar fashion as we do in this work). 
 
\end{description}

\begin{figure}
    \centering

        \includegraphics[width=1.0\columnwidth]{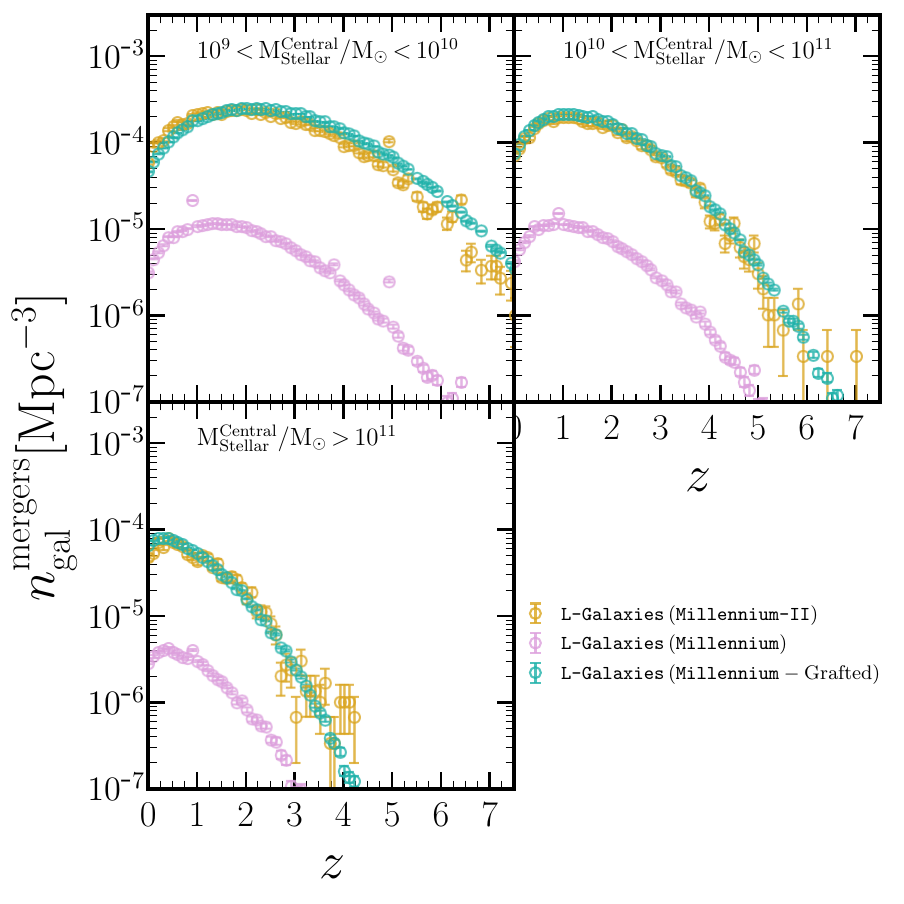}
    \caption{Number density of galaxy mergers as a function of redshift (with no cuts in mass ratios). Each panel corresponds to central galaxies with different stellar masses (from low masses to high masses starting from the top-left).  The orange symbols show the galaxy merger rate density for the \MSII,  the pink ones for the \MS{} and the green ones for the \MS{} with the \textit{grafting} procedure.}
    \label{fig:MRExt}
\end{figure}

\subsection{Effects of the \textit{grafting} on galaxy statistics}
In this work we apply the \graft{} technique included in \lgbh{}
to extend the merger trees and initialize DM halos of the \MS{} simulation.
As the high-resolution simulation, we use the \MSII{} and set the minimum value of $\delta \log_{10}(m)$ to $1.53\,{\times}\,10^{8}\,\msun/\mathit{h}$, the maximum one to $1.85\,{\times}\,10^{10}\,\msun/\mathit{h}$ and $\Delta \log_{10}(m) \,{=}\,0.1$. Thanks to this, we extended the branches of MS trees down to $\,{\sim}\,10^8\,\msun/\mathit{h}$. On the other hand, the minimum value of $\delta \log_{10}(M_h)$ is set to $1.53\,{\times}\,10^{8}\,\msun/\mathit{h}$, the maximum one to $10^{16}\,\msun/\mathit{h}$ and $\Delta \log_{10}(M_h) \,{=}\,0.1$. To guide the reader, the typical number of branches added in halos of $10^{12}\msun/\mathit{h}$ at $z\,{=}\,1$ ($z\,{=}\,0$) is $10$ ($1$).  We note that the tolerance $\Delta \log_{10}(m_h)$ is set to $0.1$ when initializing a galaxy, although we may progressively increase this value up to $0.8$ in case no matches are found (as in the case of rare massive objects at high redshift).  \\

\begin{figure}
    \centering
 
    \includegraphics[width=1.0\columnwidth]{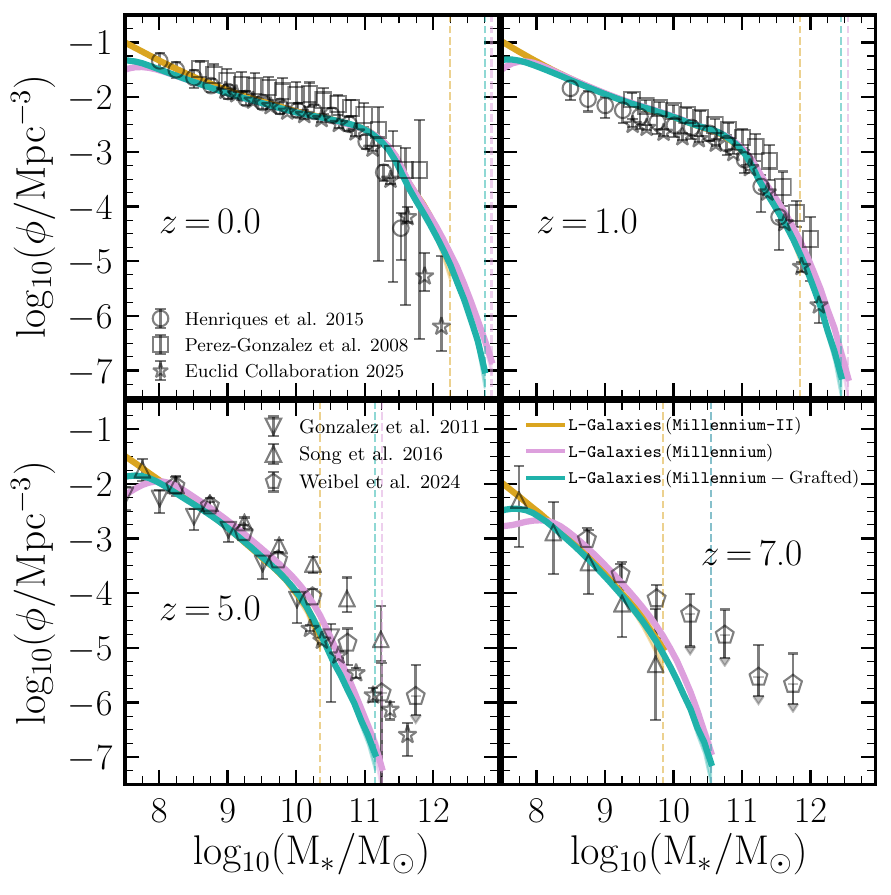}
    \caption{Stellar mass function at $z\,{=}\,0,1,5$ and $7$ predicted by \lgbh{} when run on top of the \MSII{} (orange), \MS{} (pink) and \MSG{} (green) simulations. Vertical lines highlight the maximum stellar mass found in the box. The \lgbh{} results have been compared with the observations of \cite{PerezGonzalez2008,Gonzalez2011,Song2016,Henriques2015,EuclidStellarMassFunction2025} and \cite{Weibel2024}. }
    \label{fig:SMF_MS_MSII_MSGrafted}
\end{figure}

\begin{figure}
    \centering

    \includegraphics[width=1.0\columnwidth]{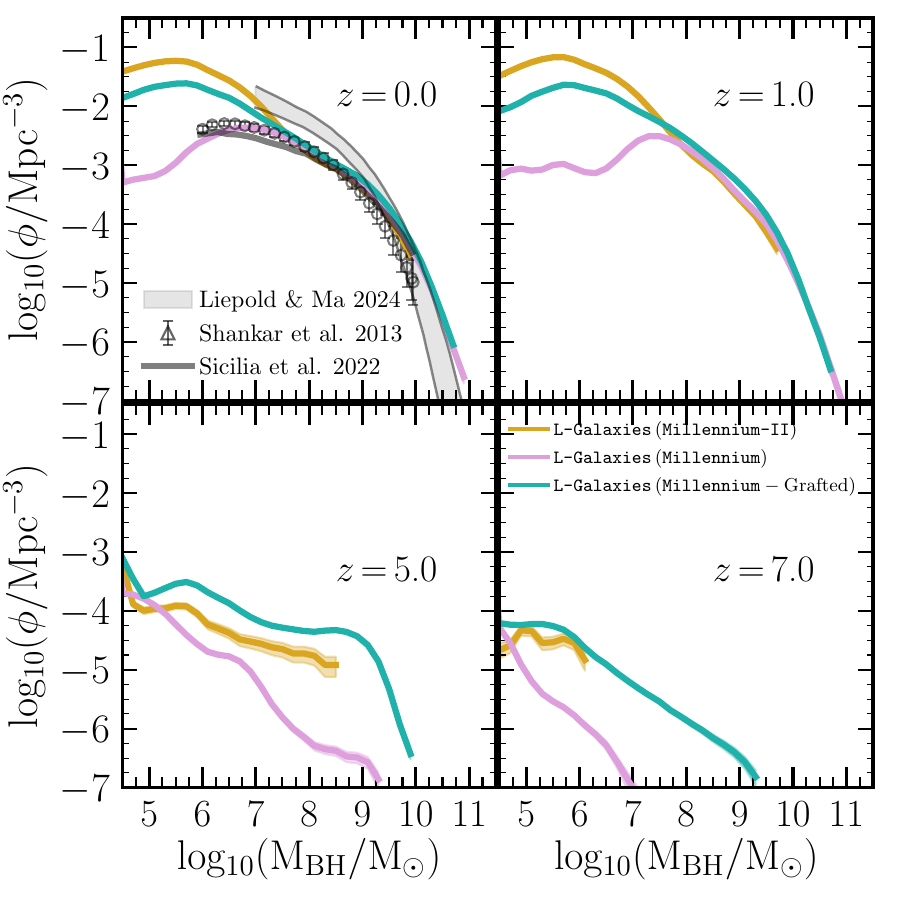}
    \caption{Black hole mass function at $z\,{=}\,0,1,5$ and $7$ predicted by \lgbh{} when run on top of the \MSII{} (orange), \MS{} (pink) and \MSG{} (green) simulations. The \lgbh{} results at $z=0$ are compared with the constraints of \cite{Shankar2013,Sicilia2022} and \cite{LiepoldAndMa2024}.} 
    \label{fig:BHMF_MS_MSII_MSGrafted}
\end{figure}

To show the potential of the \graft{} procedure, in Fig. \ref{fig:MRExt} we show the redshift evolution of the number density of galaxy mergers. We show three different stellar mass bins defined by using the mass of the central galaxy in the galaxy-merger.  We compare the results of \lgbh{} obtained on the \MS{} alone, on the \MSII{}, and on the \MSG{}. As expected, the number of mergers captured by the \MS{} alone (blue symbols) is much smaller than the one derived from the \MSII{} (orange symbols). Differences between \MS{} and \MSII{} are simply due to the higher mass-resolution of the latter, which can resolve a significant larger number of minor mergers. 
This is especially evident for the smallest galaxies (upper-left panel in Fig.~\ref{fig:MRExt}), for which the \MS{} can effectively only detect equal-mass mergers. On the other hand, the \MSII{} struggles to provide a large statistical sample of massive galaxies (bottom left panel), in particular at $z\,{\gtrsim}\,3$, where these massive objects are rare. Finally, purple symbols show the galaxy-merger number density obtained for the \MSG{}. As shown, the \graft{} procedure allows the recovery of a similar merger rate as the one predicted by the \MSII{} at all masses. Furthermore, this procedure also increases the statistics of mergers for the most massive galaxies at high redshift. More precisely,  \graft{} allows us to jointly exploit the advantages provided by the higher resolution of the \MSII{} to the better statistics offered by the larger volume of the \MS{}.\\ 

The \graft{} procedure provides similar advantages to a number of different properties simulated by \lgbh{}. As an example, in Fig.~\ref{fig:SMF_MS_MSII_MSGrafted} we show the evolution of the stellar mass function obtained for the \MS{}, \MSII{} and \MSG{}. It is evident that the results obtained for these three configurations align strikingly well on a wide range of masses and redshifts. In particular, \lgbh{} on the \MSG{} (green line) produces a similar stellar mass function as when run on the \MS{} (pink line), despite the significantly different amount of galaxy mergers of the two simulations (see Fig.~\ref{fig:MRExt}). Similarly, the results on the \MSG{} closely follow those obtained on the \MSII{} over large intervals of stellar mass and redshift. However, the \MSG{} shows improvements with respect to the \MS{} at $\mstar \,{<}\, 10^8 \, \msun$ and $z\,{\geq}\,5$. Therefore, the \graft{} allows us to improve the SMF predictions of the \MS{} for the lowest stellar mass bins and highest redshifts, without sacrificing the statistics of the most massive galaxies provided by the large volume of the \MS{}. We show this in Fig.~\ref{fig:SMF_MS_MSII_MSGrafted} by highlighting the largest $\mstar$ found in the box of each simulation with dashed vertical lines. The maximum $\mstar$ found for the \MS{} and the \MSG{} 
 are very similar, and  differ at most by $<\!0.25$dex. At the same time, they are systematically larger (by ${\gtrsim}\,0.25\,{-}\,0.75$ dex) than the largest $\mstar$ found for the \MSII{}.\\

Similarly to the stellar mass function, and before detailing all the MBH physics that \lgbh{} features, in Fig.~\ref{fig:BHMF_MS_MSII_MSGrafted} we showcase the black hole mass function (BHMF) obtained using the merger trees of the \MS{}, \MSII{}, and \MSG{}. %Analogously to what we showed in Fig.~\ref{fig:SMF_MS_MSII_MSGrafted}, we show the results at $z\,{<}\,8$,  
At $z\,{=}\,0$ all the runs of \lgbh{} across the different merger trees are consistent with each other and also align with the observed BHMF at masses $\rm {>}\,10^7 \, \msun$ \citep{Shankar2013,Sicilia2022,LiepoldAndMa2024}.  We stress that the \MSG{} extends the predictions of the \MSII{} at the high-mass end by approximately one order of magnitude. This difference is primarily due to the larger volume covered by the \MSG{} combination, which allows for better statistical sampling of the most massive and rare MBHs. Conversely, for $\rm M_{BH}\,{<}\,10^7 \, \msun$, the BHMF produced by \lgbh{} when run on \MS{} diverges from that obtained on the \MSII{} while the run on the \MSG{} shows an intermediate behaviour\footnote{The small differences observed between \MSII{} and \MSG{} at low MBH masses are not unexpected. Although combining \lgbh{} with \MSG{} enables us to generate merger trees at the same \MSII{} resolution, we are still unable to recover trees where the central halo of the main branch (i.e., the MB tag in Fig.~\ref{fig:MTE}) falls below the \MS{} resolution. As a result, some discrepancies in the BMHF at the low-mass end between \MSII{} and \MSG{} are still anticipated.}. The $z\,{=}\,1$ BHMF predicted by \lgbh{} across the three simulations, exhibits trends similar to those observed at $z\,{=}\,0$. This suggests that the main features of the BHMF are already established by then \citep[see similar results in][]{MerloniANDHeinz2008}. 
However, unlike at $z\,{=}\,0$, the number density and mass range of BHMF at $\rm M_{BH}\,{>}\,10^{7.5}\msun$ derived from the \MSG{} simulation appears to be higher than those obtained from the \MS{}. These details suggest that the galaxy merger history unresolved by the \MS{} can impact the mass-growth history of MBHs across cosmic time. This is particularly significant at high redshfit:  the $z\,{=}\,5$ and $z\,{=}\,8$ BHMFs obtained with the \MSG{} show a faster build-up of the MBH population with respect to \MS{}. Specifically, \lgbh{} applied on \MSG{} reaches MBHs as massive as $\rm M_{BH}\,{\gtrsim}\,10^7\,\msun$ already at $z\,{=}\,8$ and well above $\rm M_{BH}\,{>}\,10^9\msun$ at $z\,{=}\,5$. These mass ranges are inaccessible by either the \MS{} (because of resolution and loss of small mergers that can still drive significant amount of gas to the center) or the \MSII{} (because of volume), showing the crucial importance of the \graft{} procedure when studying populations of high-$z$ MBHs. Unfortunately, we underline that the \MS{} (and, by extension,  also the \MSG{}) does not simulate a sufficiently large volume to draw statistically reliable conclusions about the population of MBHs with $\rm M_{BH}\,{>}\,10^{9}\, \msun$ at $z\,{>}\,6$  observed by high-$z$ quasar surveys \citep[e.g.][for a recent review]{fan_banados_simcoe2023review}, but it samples accurately the assembly of the more moderate-mass black holes recently detected by JWST, as we will see in the result sections.

\subsection{The L-Galaxies model: Galaxies and MBHs}

\lgbh{} is a branch of \lgal{} specifically tailored to study the formation and evolution of MBHs. For this work, all the galaxy-evolution physics included in \lgbh{} is inherited from the version of \lgal{} presented in \cite{Henriques2015}, of which we provide a brief outline here.\\

Using as a foundation the framework proposed by \cite{WhiteandRees1978}, \lgal{} assumes that the evolution of baryons starts with the infall of gas from the inter-galactic medium within  collapsed DM halos. During this process, a hot gas atmosphere forms and gradually cools down at the centre of halos, forming disc-like structures. As soon as the cold-gas mass exceeds a certain threshold, star formation processes give rise to a stellar disc component. At the same time, supernovae explosions powered by the death of massive  stars inject energy into the interstellar medium driving the ejection of cold gas back to the hot-gas phase.  The morphological transformation of galaxies, with stellar discs evolving into bulge components, happen  through disc instability events (DI, secular processes) or galaxy mergers. The latter are driven by dynamical friction after the coalescence of the host DM halos. Finally, \lgal{} also accounts for environmental processes such as hot gas stripping during galaxy interactions. For further details, we refer the reader to \cite{Guo2011} and \cite{Henriques2015}. We stress that the  \MS{} and \MSII{} snapshots are separated by ${\approx}\,300\, \rm  Myr$. To improve the tracing of the baryonic physics involved in galaxy evolution, \lgal{} performs an internal time interpolation of 5-50 Myr (depending on redshift).

\begin{figure*}
    \centering
 
    \includegraphics[width=2\columnwidth]{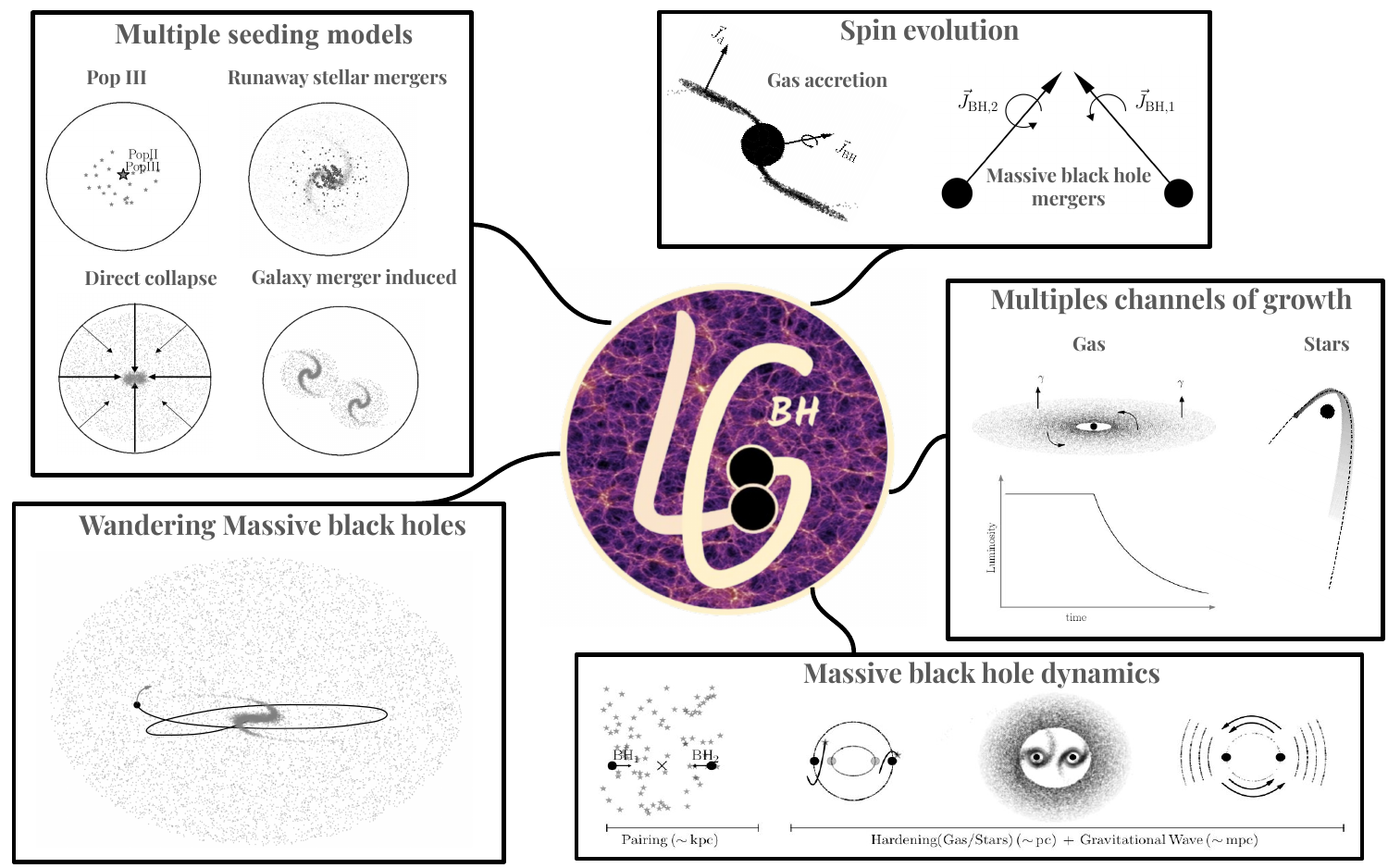}
    \caption{Illustration of the physical processes regarding MBH formation and evolution currently included in the \lgbh{} model. Clockwise from top-left: \textit{multi-flavour} seeding, spin evolution, different growth channels (from gas and tidal disruption of stars), MBH binary dynamics and formation and evolution of wandering MBHs. All these processes are tracked on-the-fly by \lgbh{}. }
    \label{fig:LGBHlogo}
\end{figure*}

\subsection{Massive black holes: formation, growth and dynamics} \label{sec:MBHs_MBHBs}

\lgbh{} includes a refined physical model to follow the genesis and evolution of MBHs and MBHBs. In this section, we outline the main physics included. A graphical summary of all the physical processes modelled in \lgbh{} is shown in Fig.~\ref{fig:LGBHlogo}. As detailed in this section, these include: a comprehensive BH-seeding model, a detailed prescription for the mass growth of MBHs via gas and stellar accretion, which is fully coupled to the physical models for MBHs spin evolution, an extensive model for the dynamics of MBHs, which tracks both the formation and evolution of MBH binaries as well as the evolution of wandering MBHs after gravitational recoils or galaxy tidal disruptions. In addition, a companion paper (Herrero-Carrión et al., in prep.) will present a detailed modelling included in \lgbh{} to determine the spectral energy distribution (SED) emerging from accreting MBHs. This includes the continuum emission produced by the accretion disc, corona, and dusty torus, as well as the emission lines originating from both the broad-line and narrow-line regions.

\subsubsection{Black hole seeds} \label{sec:seeds}

The model for the formation of MBHs in \lgbh{} is an extension of the one presented in \cite{Spinoso2022}. Here we describe its most relevant features, leaving its details to be presented by Spinoso et al (in prep). Specifically, \lgbh{} takes into account the formation of \textit{light}, \textit{intermediate}-mass and \textit{heavy} MBH seeds, according to four different channels, namely:

\begin{itemize}
\item Light PopIII remnants (i.e. light seeds, or LSs hereafter),
\item Intermediate-mass BHs formed through runaway stellar mergers (RSM) within dense stellar clusters,
\item Heavy, direct collapse BHs (DCBHs) formed in metal-free proto-galaxies,
\item Heavy seeds formed after the merger of evolved galaxies (i.e. merger-induced direct-collapse BHs, miDCBHs).
\end{itemize}
Regarding the formation of light, RSM and DCBH seeds, \lgbh{} includes a highly versatile BH seeding model. In fact, it is possible to employ either: i) physically motivated prescriptions \citep[largely based on][]{Spinoso2022} or ii) probabilistic and phenomenological prescriptions. In both cases, the formation of miDCBHs is determined only by the properties of the merging galaxies and their DM hosts, closely following the model of \cite{Bonoli2014}. Independently of their specific seeding prescription (described below), MBHs in \lgbh{} form with an initial spin value randomly selected in the $[0,1]$ range\footnote{The MBH spin evolution model is described in \ref{sec:spin}}. In addition, this work assumes that all MBH seeds form at the center of proto-galaxies. This might not be the case, in particular for the case of light seeds born in stellar groups or clusters \citep[e.g.][]{greif2012,desouza_basu2015,park_ricotti_kazuyuki2024}. Although \lgbh{} actually includes the formation of off-center MBH seeds, the analysis of this feature is beyond the scope of this work. The consequences induced by off-center MBH seeds formation onto the evolution of MBHs will be studied in detail by Izquierdo-Villalba (in prep.).\\

\noindent (i) \textbf{Physically motivated BH seeding}: As in \cite{Spinoso2022}, \lgbh{} models the occurrence of light, RSM or DCBH seeds according to the local values of the Lyman-Werner (LW) illumination and of the IGM metallicity ($\rm Z_{IGM}$) computed at the positions of each newly resolved DM halo. Both the local LW and $\rm Z_{IGM}$ fields are tracked on-the-fly as the composition of uniform backgrounds and spatial variations.  

Modeling the formation of LSs, on the other hand, is tricky in \lgbh{} because the mass resolution of the \MSII{} (i.e. the highest-resolution simulation currently available to \lgbh{}) is too coarse to directly track the evolution of PopIII stars into compact remnants. For this reason, \cite{Spinoso2022} introduced evolved LSs 
by exploiting external inputs from the \texttt{GQd} model \citep{valiante2016,Valiante2021}. Instead, \lgbh{} models the formation of LSs in a subgrid fashion, which will be fully detailed by Spinoso et al. (in prep.). In summary: only DM halos that miss the physically motivated conditions to form RSM or DCBH seeds can be initialized with evolved LSs, according to a given probability which depends on the halo mass ($\mhalo$) and redshift, following \citet{Spinoso2022} (Eq. 9.) and \cite{IzquierdoVillalba2023} (Eq. 1):
\begin{equation}
    \rm\mathcal{P}_{seed} = \mathcal{A}_{seed}\,(1+\mathit{z})^{7/2}\,\left(\frac{\mhalo}{7\times10^{10}\, \msun}\right),
    \label{eq:seeding_probability}
\end{equation}
with $\mathcal{A}_{seed}$ being a normalization constant, whose value is chosen to reproduce the occupation fraction of newly-formed BH-seeds presented in \cite{Spinoso2022}. 

As soon as a potential LS host is resolved at a given redshift $z_{\rm\,r}$, we assume that a PopIII star-formation (SF) event has happened at an \textit{earlier} redshift $z_{\rm seed}$ (extracted randomly within $z_{\rm\,r}\,{<}\,z_{\rm seed}\,{<}\,35$).

This unresolved PopIII SF event is assumed to produce a given stellar mass $\mstar^{\rm PopIII} = \alpha_{\rm SF}M_{\rm gas}$,  where $\alpha_{\rm SF}=0.025$ as in \cite{Henriques2015} and $\mathit{M}_{\rm gas}$ is the halo gas content at $z_{\rm r}$. We then sample a Larson initial mass function (see Eq.1 in \citealt{larson1998}) from $\mstar^{\rm PopIII}$, setting the IMF scale mass to $m_1=20\, {\rm \msun}$. Similarly to  \citealt{valiante2016}, the initial mass function is sampled within the mass interval [$\rm 10\,{-}\,300 \,\msun$] until $\mstar^{\rm PopIII}$ is reached. 
The most massive star obtained during the sampling is assumed to be the progenitor of the LS, whose mass is set equal to that of its stellar progenitor (typically a few $\rm{\sim}\, 10\, \msun$), hence neglecting mass losses. 
During this process, stars in the mass interval $\rm 140\,{<}\mstar/\msun\,{<}\,260$ do not form a compact remnant due to the pair-instability gap\footnote{Recent works have highlighted the possibility that single PopIII stars may produce remnants in the lower-limit of the pair-instability gap \citep[see e.g.][]{tanikawa2020,tanikawa2021,Tanikawa2024}. We will explore this possibility in upcoming works focusing on BH-seeding prescriptions.} \citep[see][]{heger_woosley2002}.

Soon after their formation, LSs can evolve in mass during phases of \textit{unresolved growth} (i.e. at $z_{\rm r}\,{<}\,{z}\,{<}\,{z_{\rm seed}}$), which proceed at the Eddington rate (i.e. $f_{\rm Edd}\,{=}\,1$), consuming cold gas from the baryonic reservoir of the newly resolved DM halo (see \S\ref{sec:growth} for details). We have checked that lowering the rate of this unresolved growth does not significantly affect our results, given the relatively small mass available for the unresolved BH growth. Finally, we stop looking for potential hosts of LSs 
%at $z\,{=}\,5.5$, 
when $\rm Z_{IGM}=10^{-3}Z_\odot$ \citep[with $Z_\odot=0.0138$, see][]{asplund2009}
as this ensures to replicate the results of \cite{Spinoso2022} and \cite{IzquierdoVillalba2023}. \\

\noindent
(ii) \textbf{Phenomenological BH seeding}: Despite the physical details of the model described in the previous paragraph, it is interesting to directly control the relative abundance of light, intermediate and heavy seeds through a parametric prescription. For instance, this allows exploring scenarios where MBHs form only through one channel and hence constraining the effects of specific BH seeding models on the global properties of MBH populations. Consequently, \lgbh{} features a phenomenological model of BH formation based only on a set of input parameters. In this model, all newly resolved DM halos have a probability to host a BH seed according to Eq. \ref{eq:seeding_probability}. The relative occurrence of light (L), intermediate (I) and heavy (H) seeds is controlled with three probability parameters, namely: $\mathcal{P_{\rm L}}$, $\mathcal{P_{\rm I}}$ and $\mathcal{P_{\rm H}}$ (with $\mathcal{P_{\rm L}}+\mathcal{P_{\rm I}}+\mathcal{P_{\rm H}}\,{=}\,1$). In other words, for each newly resolved halo that can host a BH seed (according to Eq. \ref{eq:seeding_probability}), the \textit{flavor} of the BH seed is determined randomly, according to $\mathcal{P_{\rm L}}$, $\mathcal{P_{\rm I}}$ and $\mathcal{P_{\rm H}}$. This allows to control \textit{a priori} the relative occurrence of each seed flavor. For instance, by setting $\mathcal{P_{\rm L}}\,{=}\,0.5$ and $\mathcal{P_{\rm I}} \,{=}\, \mathcal{P_{\rm H}} = 0.25$, intermediate and heavy seed types have the same probability to form, which is half of the probability for light seeds to form. Once the BH seed flavor is set, the initial seed mass is randomly extracted from Gaussian distributions determined by the free-parameters $\left\langle \mathit{M}_{\rm seed}\right\rangle$ and $\rm \sigma_{M_{seed}}$.

\subsubsection{Massive black hole growth} \label{sec:growth}
Once MBH seeds form within their host halos, 

they can start accreting gas that reaches the galaxy nuclear region or capturing close-by stars which fall within the MBH tidal radius. In the following, we describe the physical processes included in our model to trace the mass assembly of MBHs.\\

\noindent (i) \textbf{Hot gas accretion}: Following \cite{Croton2006} and \cite{Henriques2015}, the MBH can accrete part of the hot gas which surrounds the galaxy. The rate at which this happens is:
\begin{equation}\label{eq:Radio_mode}
\dot{\mathit M}_{\rm  BH} \rm \,{=}\, \mathit{k}_{AGN} \left( \frac{\mathit{M}_{hot}}{10^{11}\msun} \right) \left( \frac{\mathit{M}_{BH}}{10^{8}\msun}\right),
\end{equation}
where $\rm \mathit{M}_{hot}$ is the total mass of hot gas surrounding the galaxy and $\rm \mathit{k}_{AGN}$ is a free parameter ($1.7\,{\times}\,10^{-5}\,\rm \msun yr^{-1}$) used to reproduce the turnover at the bright end of the galaxy luminosity function. As pointed out by \cite{Marulli2008} and \cite{Bonoli2009}, the accretion rate of Eq.~\eqref{eq:Radio_mode} is orders of magnitude below the BH Eddington limit. Thus, the contribution of this accretion mode to the MBH growth is minimal. However, the AGN feedback generated during this phase (the so-called \textit{radio-mode feedback}) is essential to inject enough energy into the galaxy hot atmosphere to decrease or even stop the gas cooling rate in the galaxy \citep{Croton2006,Bower2006}.\\ 

\noindent (ii) \textbf{Cold gas accretion}:  Galaxy mergers and disc instabilities are the main processes bringing gas to MBHs and triggering their growth. Following \cite{KauffmannandHaehnelt2000}, \lgbh{} assumes that after a galaxy merger, the
gas mass reaching the center and made available for accretion is:\begin{equation}\label{eq:QuasarMode_Merger}
\rm   \Delta {\mathit M}_{BH}^{gas} \,{=}\, \mathit{f}_{BH}^{merger} (1+\mathit{z}_{merger})^{\alpha} \frac{\mathit{m}_{sat}/\mathit{m}_{cen}}{1 + (\mathit{V}_{BH}/\mathit{V}_{200})^2}\, \mathit{M}_{\rm gas},
\end{equation}
where $\rm \mathit{m}_{sat}/\mathit{m}_{cent}\,{\leq}\,1$ is the ratio between the baryonic masses of the satellite and the central galaxy, $\rm \mathit{V}_{200}$ is the virial velocity of the DM subhalo hosting the central galaxy, $z_{\rm merger}$ is the galaxy merger redshift and $\mathit{M}_{\rm gas}$ is the cold gas mass of the remnant galaxy. $\rm \mathit{f}_{BH}^{\rm merger}$ ($0.028$), $\rm \mathit{V}_{BH}$ ($\rm 280 \, km/s$) and $\alpha$ ($5/2$) are free parameters, tuned to match the black hole mass function and the $z\,{<}\,2$ quasar luminosity function. 
The $(1+z)^{\alpha}$ dependency accounts for the fact that high-$z$ galaxies are more compact than their low-$z$ analogous \citep[][]{Mo1998,Shen2003,vanderWel2014,Lange2015}, hence that gas is brought onto MBHs more efficiently at high-$z$ than at low-$z$ \citep[see also][]{Bonoli2009,IzquierdoVillalba2020,IzquierdoVillalba2021}. \\

Similarly, during phases of disc instability in the galaxy, we assume that a fraction of the galaxy gas content gets torqued towards the center \citep[e.g.,][and references therein]{Fanali2015}. In this case, the maximum mass that can be accreted by the MBH is \citep[see][]{IzquierdoVillalba2020}:
\begin{equation}\label{eq:QuasarMode_DI}
\rm    \Delta {\mathit M}_{BH}^{gas} \,{=}\, \mathit{f}_{BH}^{DI} (1+\mathit{z}_{DI})^{\alpha} \frac{\Delta \mathit{M}_{stars}^{DI}}{{1 + (\mathit{V}_{BH}/\mathit{V}_{200})^2}},
\end{equation}
where $\rm \Delta \mathit{M}_{stars}^{DI}$ represents the amount of stellar mass from the galactic disc that needs to be transferred to the bulge to recover the disc stability\footnote{Disc instabilities implemented in \lgbh{} are based on the \cite{Efstathio1982} criterion which establishes that the amount of matter which triggers a disc instability event is set to $\rm \Delta \mathit{M}_{stars}^{DI} \,{=}\, \mathit{M}_{\rm \star, d}\,{-}\, \left( \mathit{V}_{max}^2 \mathit{R}_{\star,d}/G \epsilon^{2}\right)$. The variable $\epsilon$ is a free parameter set to $1.5$, $\rm \mathit{V}_{max}$ $\rm \mathit{R}_{\star,d}$ and  $\rm \mathit{M}_{\star,d}$ correspond to the maximum circular velocity of the host DM, the scale-length and stellar mass of the stellar disc, respectively.}. $\rm \mathit{z}_{DI}$ is the redshift at which the disc instability takes place. As in Eq. \ref{eq:QuasarMode_Merger}, $\rm \mathit{f}_{BH}^{DI}$,  $\rm \mathit{V}_{BH}$, and $\alpha$ are free parameters. $\rm V_{BH}$ and $\alpha$ have the same value as in the case of mergers, while $\rm \mathit{f}_{BH}^{DI}$ ($0.0015$) assumes lower values with respect to  $\rm \mathit{f}_{BH}^{\rm merger}$ as the efficiency of driving gas to the nuclear region is likely smaller during disk instabilities than during galaxy mergers.\\

After a galaxy merger or disc instability, the cold gas available for accretion, i.e. $\rm \Delta \mathit{M}_{BH}^{gas}$, from Eq.~\eqref{eq:QuasarMode_Merger}  and/or Eq.~\eqref{eq:QuasarMode_DI} is added to a reservoir around the black hole, $\rm M_{Res}$\footnote{Notice that $\rm \mathit{M}_{Res} \,{=}\Delta \mathit{M}_{BH}^{gas}$ is only satisfied if before the galaxy merger of disc instability the reservoir around the MBH was empty. On the contrary, $\rm \mathit{M}_{Res} \,{=}\, \Delta \mathit{M}_{BH}^{gas}\,{+} \mathit{M}_{gas}^{left-over}$ being $\rm \mathit{M}_{gas}^{left-over}$ the leftover gas inside the reservoir, accumulated trough prior mergers or disc instabilities and not consumed by the MBH by the time at which the new merger or disc instability takes place.}. Instead of being instantaneously consumed, the gas reservoir is progressively exhausted in time as MBH masses evolve as:
\begin{equation}
    \rm \frac{\mathit{d}\,\mathit{M}_{BH}(t)}{\mathit{dt}} = \mathit{f}_{Edd}(\mathit{t})\,\frac{1-\eta(\mathit{t})}{\varepsilon(\mathit{t})}\,\frac{\mathit{M}_{BH}(\mathit{t})}{\tau_{\,Edd}}
    \label{eq:BH_mass_evolution}
\end{equation}
with $\rm\tau_{\,Edd}\,{=}\,\sigma_T/(4\pi G\,\mathit{M}_{BH}\,\mathit{m}_p\,c)\,{=}\,0.45\,Gyr$, where $\sigma_T$ is the Thomson scattering cross-section and $m_p$ is the proton mass. The $\eta$ and $\varepsilon$ parameters in Eq.~\ref{eq:BH_mass_evolution} respectively are the accretion and radiative efficiencies. As better detailed in Section~\ref{sec:spin}, \lgbh{} couples the value of $\eta(t)$ to the MBH-spin time evolution. On the other hand, $\varepsilon$ relates to the state of the accretion flow onto the MBH: $\varepsilon=\eta$ for thin-disk accretion \citep{ShakuraSunyaev1973} while for advection-dominated accretion flows \citep[ADAF, ][]{Rees1982} the value of $\varepsilon$ follows the prescription detailed in \cite{MerloniANDHeinz2008} \citep[see also][ for further details on the implementation in \lgbh{}]{IzquierdoVillalba2020}.\\

The Eddington rate parameter $\rm\mathit{f}_{Edd}\,{=}\,\mathit{L}_{bol}/\mathit{L}_{Edd}$ is defined in terms of the MBH bolometric luminosity ($\rm L_{bol}$) and the Eddington luminosity $\rm \mathit{L}_{Edd}\,{=}\,(4\pi G\,\mathit{M}_{BH}\,\mathit{m}_p\,c)/\sigma_T$. \lgbh{} uses $\rm\mathit{f}_{Edd}$ to modulate in time the rate of mass-accretion onto BHs through a given \textit{lightcurve}, i.e. $\rm\mathit{f}_{Edd}(\mathit{t})$. In this work, we follow \cite{IzquierdoVillalba2024} which assumed two different lightcurves depending on the environment in which the MBH is embedded and the properties of the gas inflow. The environment of the MBH is evaluated according to the mass ratio between the gas reservoir and the MBH:
\begin{equation}
\rm \mathcal{R}\,{=}\, \frac{\mathit{M}_{Res}(\mathit{t}_0)}{\mathit{M}_{BH}(\mathit{t}_0)},   
\label{eq:Mreservoir_over_MBH}
\end{equation}
and the strength of the gas inflow parameterized as:
\begin{equation}
\rm \dot{\mathit{M}}_{inflow} = \frac{\Delta \mathit{M}_{BH}^{gas}}{\mathit{t}_{dyn}}\ .
\label{eq:Minflow}
\end{equation}
where $t_0$ in Eq.~\ref{eq:Mreservoir_over_MBH} denotes the time at which the merger or disc instability takes place and $t_{\rm dyn}$ in Eq.~\ref{eq:Minflow} is the dynamical time of the galactic disc at $t_0$. \lgbh{} sets $\rm \mathit{t}_{dyn}=\mathit{V}_{disc}/\mathit{R}_{gas}^{sl}$, being $\rm \mathit{V}_{disc}$ the maximum circular velocity of the cold gas (assuming an exponential disk profile) and $R_{\rm gas}^{\rm sl}$ the cold gas scale length radius \citep[see][for further details about the disc radius model]{Guo2011}. Following \cite{IzquierdoVillalba2024}, the specific combination between $\mathcal{R}$ and $\rm \dot{\mathit{M}}_{inflow}$ determines if the merger or disc instability event created the favorable conditions to prompt super-Eddington growth or a regular Eddington-limited growth \citep[see e.g][]{Inayoshi2016,Takeo2018,Regan2019,Toyouchi2021,Sassano2023,Massonneau2023}. According to \cite{IzquierdoVillalba2024} we set $\mathcal{R}^{\rm th}\,{=}\,2\,{\times}\,10^3$ and $\rm \dot{\mathit{M}}_{inflow}^{\rm th}\,{=}\,10 \, \rm \msun/yr$ as the minimum values to trigger super-Eddington events inside \lgbh{}. The super-Eddington and Eddington-limited lightcurves used for these phases are\footnote{We stress that if a galaxy undergoes a new merger or DI while the central MBH is still accreting mass from a previous event, the new cold gas driven around the MBH environment is added to the previous remnant $\rm \mathit{M}_{Res}$ and the conditions to trigger one or another lightcurve are re-evaluated.}:

\begin{itemize}
    \item The \textit{Eddington-limited} case: Following the results from the hydrodynamical simulations of \cite{Hopkins2005}, this model for the  lightcurve associated to each accretion episode, assumes an initial Eddington-limited growth phase followed by a phase of low accretion rates \citep[see also][]{Marulli2006,Bonoli2009}, namely:
    \begin{equation} \label{eq:feed_Edd}
        f_{\rm Edd} (t) \rm \,{=}\, \left \{ \begin{matrix}
         1 & \mathrm{\mathit{M}_{BH}}(t) \, {\le}\, \mathrm{\mathit{M}_{E}}   \\ \\
         \frac{1}{\left[ 1 + ((\mathit{t}-t_0)/\mathit{t}_\mathit{Q})^{1/2}\right]^{2/\beta} } &  \mathrm{\mathit{M}_{BH}}(t) \, {>} \, \mathrm{\mathit{M}_{E}}   \\ \\
         \end{matrix}\right.
    \end{equation}
where $\mathrm{\mathit{M}_{E}}\,{=}\, \mathrm{\mathit{M}_{BH}}(t_0) + \mathcal{F}_{\rm Edd} \mathrm{\mathit{M}_{Res}}(t_0)$ sets the duration of the Eddington limited phase and it corresponds to the mass reached by the MBH after consuming a fraction $\mathcal{F}_{\rm Edd}$ of its gas reservoir \citep[with $\mathcal{F}_{\rm Edd}=0.7$ as in][]{Bonoli2009}. $\mathrm{\mathit{M}_{BH}} (t_0)$ and $\mathrm{\mathit{M}_{Res}} (t_0)$ correspond to the mass of the MBH and the reservoir at the moment of the (major/minor) merger and/or disc instability. $t_Q$ gives the time-scale at which $f_{\rm Edd}$ decreases and is defined as $ t_Q \,{=}\, t_d\,\xi^{\beta}/(\beta \ln 10)$, with $t_d \,{=}\, 1.26{\times}10^8 \, \rm yr $, $\beta \,{=}\, 0.4$ and $\xi \,{=}\, 0.3$. The specific value of these variables are based on \cite{Hopkins2009} which showed that models of \textit{self-regulated} MBH growth require $0.3\,{<}\,\beta\,{<}\,0.8$ and $0.2\,{<}\,\xi\,{<}\,0.4$.

    \item The \textit{Super-Eddington} case: During a super-critical accretion event, the lightcurve is characterized by the following $f_{\rm Edd}$:
    \begin{equation} \label{eq:feed_SE}
        f_{\rm Edd} (t) \rm \,{=}\, \left \{ \begin{matrix}
        B(\chi) [ \frac{0.985}{\dot{\rm M}_{\rm Edd}/\dot{\rm M} \,{+}\, C(\chi)}  & \\
          \,\,\, \,\,\, \,\,\, \,\,\, \,\,\, \,\,\, +  \frac{0.015}{\dot{\rm M}_{\rm Edd}/\dot{\rm M} \,{+}\, D(\chi)} ] & \mathrm{M_{BH}}(t) {\le} \mathrm{M_{SE}}\ , \\ \\
        \frac{1}{\left[ 1 + ((\mathit{t}-t_0)/\mathit{t}_\mathit{Q})^{1/2}\right]^{2/\beta} } &  \mathrm{M_{BH}}(t) {>} \mathrm{M_{SE}}\ ,   \\ \\
        \end{matrix}\right.
    \end{equation}
\noindent where the two phases capture the fact that the strong energetic feedback exerted by the super-critical gas accretion onto the inflowing gas can sweep the gas around the MBH. This would force a sub-Eddington phase within a few Myr, effectively self-regulating the MBH accretion \citep[see e.g][]{Lupi2016,Regan2019,Massonneau2023}. The parameter $\rm M_{SE} \,{=} \, \mathrm{M_{BH}} (t_0) + \mathcal{F}_{\rm SE} \mathrm{M_{Res}} (t_0)$ sets the maximum mass reached by the MBH during the super-critical accretion, with $\mathrm{M_{BH}} (t_0)$ and $\mathrm{M_{Res}} (t_0)$ being the mass of the MBH and the reservoir at the moment of the (major/minor) merger and/or disc instability. $\mathcal{F}_{\rm SE} $ specifies the fraction of $\mathrm{M_{Res}} (t_0)$ consumed by the super-Eddington phase, before the sub-Eddington one is initiated. Unlike \cite{IzquierdoVillalba2024}, we assume that $\mathcal{F}_{\rm SE}$ is not fixed but depends on $\mathrm{M_{BH}} (t_0)$. This assumption accounts for the fact that the larger the mass of the MBH the larger is the feedback injected into its surroundings, hindering prolonged super-Eddington accretion. In particular, $\log_{10}\left(\mathcal{F}_{\rm SE}\right) \,{=}\, -0.27\,\log_{10}\!\left(\rm M_{BH}/\msun\right)\,{-}\,0.72$, ensuring that the bright end of the LFs at $z\,{>}\,6$ are not over-predicted. The functions $B(\chi)$, $C(\chi)$ and $D(\chi)$ are taken from \cite{Madau2014} and they scale with the spin of the MBH ($\chi$, see \S~\ref{sec:spin}). Finally, $\dot{\rm M}$ and $\dot{\rm M}_{\rm Edd}$ are the accretion rate and Eddington accretion rate onto the MBH, respectively. To determine $\dot{\rm M}$ we extract a random number between [0-$10^5$] distributed according to $\dot{\rm M}^{-1}$ \citep[see similar power-laws in][]{Fanidakis2012,Griffin2018,Shirakata2019}. We emphasize that this super-Eddington accretion model has been previously tested and explored in \citet{IzquierdoVillalba2024}, where it was shown that episodes of super-Eddington growth in high-redshift MBHs can help alleviate the current tension between recent PTA observations and predictions from galaxy formation models. 
\end{itemize}

\noindent (iii) \textbf{Tidal disruption events and star accretion}: \lgbh{} also includes an additional module where we model the frequency of the disruption of stars reaching the vicinity of MBHs (Tidal Disruption Events, TDEs) and subsequent events of accretion of the disrupted stars. Assuming that galaxies host nuclear star clusters based on the occupation fraction of \cite{Hoyer21}, at any given time of the \lgbh{} run, we estimate the time-evolving TDE rate based on the properties of the nuclear star clusters, the galaxy stellar content and the MBH mass.  The event rate is given by tabulated values derived from  runs previously carried out with \texttt{PhaseFlow} \citep[][]{Vasiliev17, Bortolas22}. A full description of the modeling of TDEs in \lgbh{} can be found in \cite{Polkas2024}. For simplicity, we do not include this growth channel in this work, given that it contributes negligibly to the growth of MBHs, especially in the intermediate-high mass range (Polkas et al., in prep. ). \\

\subsubsection{Spin evolution: Gas and merger interplay}  \label{sec:spin}

\lgbh{} also tracks the spin ($\chi$) evolution of MBHs. As described above, at formation time, the initial MBH spin is assumed to be randomly distributed between $\left[0,0.998\right]$\footnote{The upper value refers to the Thorne limit. Radiation emitted by the disc and captured by the hole produces a counteracting torque which prevents spin-up beyond the value of 0.998 \citep{Thorne1974}.}. Subsequently, gas accretion events and MBH mergers lead to changes in the spin magnitude. In the following, we summarize the main features included by \lgbh{} to trace the spin evolution\footnote{We neglect here the secular change in the black hole spin direction in warped discs as it does not affect the spin magnitude and black hole growth \citep{Dotti2013}.}.
\\

\noindent (i) \textbf{Spin evolution due to gas accretion}: During an accretion event, the gas settles in a disc which may not lie in the equatorial plane of the rotating MBH. When this happens, there is a misalignment between the disc angular momentum (\Jdv) and the MBH angular momentum (\Jbhv). As a result, the orbital plane of the inner parts of the disc aligns (\textit{prograde} orbit, spinning up the MBH) or counter-aligns (\textit{retrograde} orbit, spinning down the MBH) with respect to \Jbhv \citep{Bardeen1975}.  \cite{King2005} demonstrated that  during an accretion event the fraction of transient discs consumed in a prograde accretion event, $n_{\rm Pa}$, is always $1$ when $|\vec{J_{\rm d}}|\,{>}\,2|\vec{J}_{\rm BH}|$ \citep{King2005}. On the contrary, when $|\vec{J_{\rm d}}|\,{<}\,2|\vec{J}_{\rm BH}|$, the value of $n_{\rm Pa}$ is determined according to \cite{Sesana2014}:
\begin{equation} \label{eq:npa}
n_{\rm Pa} = F \,{+}\, \frac{|\vec{J_{\rm d}}|}{2|\vec{J}_{\rm BH}|}(1\,{-}\,F) \, ,  
\end{equation}
where $F$ is an \textit{isotropy parameter} and it is linked with the bulge morphology \citep[see][for further details]{IzquierdoVillalba2020}. Taking into account the above relation, since an accretion event is composed by a fraction of $n_{\rm Pa}$ prograde and $(1-n_{\rm Pa})$ retrograde orbits, the rate of the spin change can be computed as \citep{Sesana2014,Barausse2012}:
\begin{eqnarray}
    \dot{\chi} &{=}& \left[ \, \left(n_{\rm Pa}\, L_{\rm ISCO}^{\rm pro}(\chi) + (1{-}n_{\rm Pa})L_{\rm ISCO}^{\rm retro}(\chi)\right)\right . \nonumber\\
 & {-} &\left . 2a\left( n_{\rm Pa}\, E_{\rm ISCO}^{\rm pro}(\chi) + (1{-}n_{\rm Pa})E_{\rm ISCO}^{\rm retro}(\chi)\right) \right]\, \frac{ \dot{\mathit M}_{\rm BH}}{\mathit M_{\rm BH}}, 
\end{eqnarray}
where $\rm \dot{M}_{BH}$ and $\rm M_{BH}$ correspond to the accretion rate and mass of the MBH, respectively. $L_{\rm ISCO}^{\rm pro}$ ($L_{\rm ISCO}^{\rm retro}$) and $E_{\rm ISCO}^{\rm pro}$ ($E_{\rm ISCO}^{\rm retro}$) are, respectively, the specific angular momentum and energy in a prograde (retrograde) black hole Innermost Stable Circular Orbit \citep[ISCO,][]{Barde1972}. Taking this into account, the value of the mass-accretion efficiency, $\eta$, after a fraction of prograde $n_{Pa}$ and ($1-n_{Pa}$) of retrograde orbits is: 
\begin{equation}\label{eq:eta_Sesana}
\eta(\chi) \,{=}\, n_{\rm Pa}\,\eta_{\rm pro}(\chi) + (1-n_{\rm Pa})\,\eta_{\rm retro}(\chi) \, ,
\end{equation}
where $\eta_{\rm pro}$ and $\eta_{\rm retro}$ are the values of accretion efficiency assuming the radius of the last stable circular orbit, $r_{\rm ISCO}$, in a prograde and retrograde orbit, respectively.\\
    
\noindent (ii) \textbf{Spin evolution due to MBH coalescence}: After two MBHs merge, the final spin of the remnant black hole  ($\chi_f$) is determined according to the law of angular momentum conservation  \cite{BarausseANDRezzolla2009}. The value of $\chi_f$ depends on the binary mass ratio, the spin magnitude of the MBHs constituting the binary system, the angle between the two MBH spins, and the angle between the MBH spins and the orbital angular momentum. Since \lgbh{} only tracks the evolution of the MBH spin modulus, the orientation between the spins and orbital angular momentum is unknown. To address this limitation, \lgbh{} distinguishes between two types of mergers \citep[see also ][]{Barausse2012,Volonteri2013}: \textit{wet} and \textit{dry}. The wet category refers to the cases where the total mass of the binary is smaller than the gas reservoir available for accretion. In these circumstances, it has been shown that the spin of the two MBHs aligns within a residual offset of 10 deg \citep{Dotti2010}. In contrast, for dry mergers, \lgbh{} assumes a random orientation between the two MBH spins.

With the inclusion of the described spin model, \lgbh{} predicts that MBHs with  $\rm M_{BH}\,{<}\,10^{6}\, \msun$ tend to be maximally spinning, while at MBHs with larger masses display lower spin values. That is, the average spin values decrease with increasing MBH mass  \citep[see][for further details]{IzquierdoVillalba2020}. This trend reflects the combined effects of coherent and incoherent gas accretion, driven by secular processes and galaxy mergers, as well as MBH coalescence. Although observational constraints on MBH spins remain limited, \cite{Reynolds2013} provides a compilation of MBHs with robust measurements of both spin and mass. As demonstrated by \cite{IzquierdoVillalba2020}, comparison of these observational data with predictions for active MBHs from \lgbh{} shows good agreement, underscoring the strong predictive power of our semi-analytical model.

\subsubsection{Massive black hole binaries: Dynamical evolution} 

Unlike other models, \lgbh{} traces the dynamical formation and evolution of MBHBs after a galaxy-galaxy merger \citep{IzquierdoVillalba2021}. In particular, it follows their evolution through the three different phases of MBH binary evolution \citep{Begelman1980}: pairing, hardening, merger.  After am unequal galaxy-galaxy merger, the MBH originally hosted by the less massive, satellite galaxy is deposited far away (i.e., at a few $\rm {\sim}\,kpc$) from the galactic nucleus of its new host. At these distances, the MBH is assumed to undergo a dynamical friction phase (modelled according to \citealt{BinneyTremaine2008}) that makes it sink towards the galactic nucleus. Once the dynamical friction phase ends, the satellite MBH reaches the primary MBH placed at the galactic nucleus, forming a gravitationally bound MBHB. At this moment, the hardening and GW evolutionary phase starts and the binary separation and eccentricity are evolved self-consistently by \lgbh{} via numerical integration of a suite of equations, depending on the environment in which the binary is placed \citep[see][]{IzquierdoVillalba2021}. In gas-rich surroundings, the interaction with a circumbinary gaseous disc drives the evolution of the system \citep{Dotti2015}. Otherwise, interaction with single stars distributed as a Sérsic density profile is responsible for the dynamical evolution of the MBHB \citep{Quinlan1997,Sesana2015}. Finally, the emission of GWs hardens the MBHB during its final evolutionary stage \citep{PetersAndMathews1963}. For further details of the single and binary MBH model, we refer the reader to \cite{IzquierdoVillalba2021}.  We stress that in the case of repeated galaxy mergers, a third MBH can reach the nucleus of the remnant before the already existing MBHB completes its evolution. In this case, an MBH triplet forms and the outcome of the triple interaction is modeled by \lgbh{} according to the tabulated values presented in \cite{Bonetti2018ModelGrid}. \\

Besides tracing the dynamical evolution of MBH pairs in galaxy mergers, \lgbh{} allows MBHs to grow during the dynamical friction and hardening phase. Modelling this type of growth is fundamental to explore dual AGNs or perform multimessenger studies. Regarding the MBHs in the dynamical friction phase, recent hydrodynamical simulations of merging galaxies have shown that the secondary galaxy undergoes large perturbations during the pericenter passages around the central one \citep[see e.g.,][]{Callegari2009,Callegari2011a,Callegari2011b,Capelo2015}. This leads to the MBH in the secondary galaxy to experience enhanced accretion events, primarily correlated with the galaxy mass ratio. \lgbh{} includes these findings by assuming that before the galaxy merger, the MBH is able to increase its gas reservoir according to Eq.~\ref{eq:QuasarMode_Merger}. Once the satellite MBH is deposited in the new galaxy and starts its dynamical friction phase, the gas reservoir is progressively consumed according to Eq.~\ref{eq:feed_Edd}. This accretion phase lasts until the MBH consumes the total gas reservoir stored before the merger. Concerning the growth of hard binaries, \lgbh{} assumes the so-called \textit{preferential accretion} found in numerous hydrodynamical simulations \citep[see e.g., ][]{DOrazio2013,Farris2014,Moody2019,Munoz2019,DOrazio2021}. In brief, the accretion rate of a primary black hole ($\dot{\rm M}_{\rm BH_1}$) is fully determined by the binary mass ratio ($q$) and the accretion rate of the secondary black hole ($\dot{\rm M}_{\rm BH_2}$) is:
\begin{equation} \label{eq:Relation_accretion_hard_binary_blac_hole}
\dot{\rm M}_{\rm BH_1} =  \dot{\rm M}_{\rm BH_2} (0.1+0.9\mathit{q}).
\end{equation}
Except in the case of equal mass systems, the secondary MBHs are farther away from the binary centre of mass than primary ones. This allows them to be closer to the circumbinary disc edges and display high accretion rates. Based on this, \lgbh{} fix the accretion of the secondary MBH at the Eddington limit and determine the accretion onto the primary according to Eq.~\ref{eq:Relation_accretion_hard_binary_blac_hole}.

\subsubsection{Wandering MBHs: Galaxy disruption and gravitational recoil}

All the MBHs in \lgbh{} are born inside the nucleus of a the host galaxy. However, many different mechanisms can move them away from their birthplaces and put them in bound orbits within the DM halo. These are called \textit{wandering} MBH (wMBH) and \lgbh{} is the only SAM to date which traces the origin and orbits of these objects. In the following paragraphs, we briefly describe the formation mechanism of wMBHs and the method used to trace their orbits.  \\

\noindent (i) \textbf{Disrupted wMBHs}: As soon as two DM halos merge, their galaxies do so  as well on a timescale given by the dynamical friction presented in \cite{Guo2011} and \cite{Henriques2015}. During this process, \lgbh{} tracks the trajectory of the smallest galaxy and checks at each time step whether it can be disrupted due to tidal forces\footnote{To determine a disruption process, \lgbh{} compares the galaxy baryonic (cold gas and stellar gas) density within the half-mass radius to the DM density of the massive halo. If the former is smaller than the latter one, the galaxy is tidally disrupted.} before merging with the central galaxy. If a disruption occurs, the satellite MBH is deposited in the DM halo as a wMBH and \lgbh{} follows its orbit by solving self-consistently its equation of motion. To do so, the code uses the final position and velocity of the galaxy before being disrupted as a starting point and accounts for the gravitational acceleration and the dynamical friction exerted by any of the inter-cluster components, such as DM, stars, and hot gas. \\

\noindent (ii) \textbf{Recoiled wMBHs}: During the final phase of MBHB evolution, the binary  reduces its separation due to the emission of GWs. At the moment of merger, this emission imparts a recoil on the remnant MBH whose velocity has been shown to be characterized by the magnitude and orientation of the MBH spins. \lgbh{} models this kick velocity using the formula established by \cite{Lousto2012}. If the magnitude of the kick velocity exceeds the escape velocity of the host galaxy, the MBH is ejected and deposited inside the DM halo. When this occurs, \lgbh{} follows the orbit of the newly formed wMBH in the same way as described for the \textit{disrupted} wMBHs.\\

We stress that as soon as a galaxy falls inside a larger DM halo, \lgbh{} assumes that it cannot retain any of its wMBHs. Instead, these start wandering into the halo of the most massive galaxy. Additionally, \lgbh{} considers that a wMBH is re-incorporated inside the galaxy when its distance from the galaxy is less than the galaxy size and its velocity is lower than the galaxy escape velocity at that distance\footnote{To compute the escape velocity at a given distance, \lgbh{} uses a \cite{Hernquist1990} profile for the galactic bulge and an exponential profile for the stellar and cold gas disc.}.  \cite{IzquierdoVillalba2020} provided an overview of the wMBH population predicted by \lgbh{}, showing that this population is approximately 2 dex less common than the nuclear MBH population, regardless of mass and redshift. wMBHs are more frequently found in more massive galaxies and tend to be located at distances between 0.1 and 0.7 times the subhalo virial radius. Additionally, \cite{IzquierdoVillalba2020} demonstrated that incorporating the modelling of wMBH populations introduces a distinctive signature in the $z=0$ galaxy–MBH scaling relations, increasing their scatter and breaking the correlation in massive galaxies that host pseudobulge structures (as these galaxies are unable to retain  MBHs when they are recoiled).  We refer the reader to \cite{IzquierdoVillalba2020} more details on the modeling and global properties of the wMBH population in \lgbh{} and \cite{Untzaga2024} for the study of the wMBHs expected in Milky Way-type galaxies.

\section{Physics variations of \lgbh{} used in this work}\label{sec:ALL_model_variations}

\begin{figure}
    \centering
    \includegraphics[width=1.0\columnwidth]{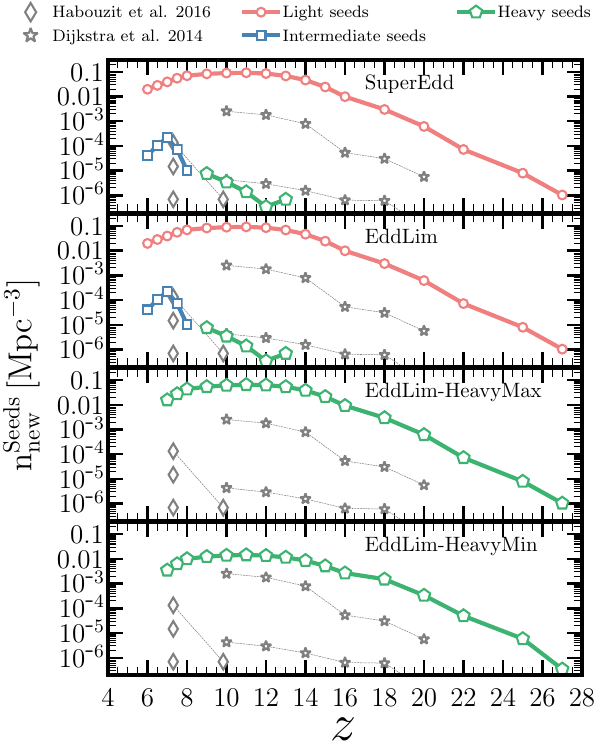}
    \caption{Number density of the newly formed seeds predicted by \lgbh{} when run on the \MSII{} merger trees: Light (red), Intermediate (blue) and heavy Heavy (green) seeds. Each panel shows one of the models explored in this work: \supereddington{}, \eddington{}, \heavyMax{} and \heavyMin{}. The results are compared with the predictions for heavy seed formation of \cite{Habouzit2016} (diamonds) and two of the model variations presented in \cite{Dijkstra2014} (stars). In particular, the highest of these two latter data series shows the case in which metallic feedback from SNe is neglected, resulting in an artificially high number density of heavy seeds.}
    \label{fig:Newly_Formed_Seed_All_Models}
\end{figure}

We present here the \lgbh{} configurations adopted to investigate the physical processes required for MBHs to simultaneously reproduce PTA (low-$z$) and JWST (high-$z$
observations.  Four model variations are explored: two of them are linked to the rate at which MBHs grow over cosmic time (\supereddington{} and \eddington) while the other two  investigate the possibility that MBHs form exclusively as heavy seeds (\heavyMax{} and \heavyMin{}).  All models include spin evolution, MBHB dynamical evolution and formation of wandering MBHs through GW recoils and galaxy disruptions. Unless otherwise stated, the results presented for all models are for \lgbh{} run on the \MSG{}.  The four models are:

\begin{itemize}
    \item[\ding{109}]  \textbf{\supereddington}: This model includes the physically motivated MBH seeding described in the first part of Sect.~\ref{sec:seeds}  \citep[i.e., the one presented in][]{Spinoso2022}. On top of mass accretion at  Eddington and sub-Eddington rates, it also allows for phases of Super-Eddington accretion, as described in Sec.~\ref{sec:growth}. \\ 

    \item[\ding{109}] \textbf{\eddington}: This model version is identical to the \supereddington{} one, except that MBHs cannot undergo super-Eddington accretion regardless of the gas reservoir content and the gas inflow rate after galaxy mergers or disc instabilities (i.e., growth is always capped at the Eddington limit).\\ 
    
    \item[\ding{109}]  \textbf{\heavyMax}: In this model we modify the MBH seeding prescription. Instead of using the physically motivated model used in the first two models, it uses the phenomenological seeding prescription described in the second part of Sect.~\ref{sec:seeds}, with $\mathcal{P_{\rm L}} \,{=}\, \mathcal{P_{\rm I}} =0$ and $\mathcal{P_{\rm H}}\,{=}\,1$. This implies that only heavy seeds can form. As for the initial distribution of seed masses, we set $\rm\left\langle M_{seed}\right\rangle=5\times10^4\msun$ and $\rm\sigma_{M_{seed}}\,{=}\,0.15$ (see Sect.~\ref{sec:seeds}), in line with current hypotheses about heavy seed masses \citep[e.g.][]{lodato_natarajan2006, Woods2019}. For this model realization we set $\rm\mathcal{A}_{seed}=0.02$ in Eq.~\ref{eq:seeding_probability}. This value is chosen to produce an {\it artificially} high number-density of  heavy seeds, i.e. 
    the same number density as 
    the one we obtain for light seeds within our physically-motivated seeding model (see Fig.~\ref{fig:Newly_Formed_Seed_All_Models}).  This assumption lacks a physical justification within our current understanding of seed formation mechanisms. Heavy seeds are permitted to form as early as light seeds, thus  establishing an upper limit on both the formation time and occupation fraction of $\sim 10^4\msun$ seeds.  This model has been developed  to assess whether heavy seeds alone could account for both the strong sGWB signals reported by PTA experiments and the recent JWST findings. In this model growth is capped at the Eddington limit, as for model \eddington{}.\\
   
    \item[\ding{109}]  \textbf{\heavyMin}: This is identical to the \heavyMax{} model except for a less-efficient formation of heavy seeds. In detail, we keep $\mathcal{P_{\rm L}} = \mathcal{P_{\rm I}} =0$ and $\mathcal{P_{\rm H}}=1$ so that only heavy seeds can form, with the same mass distribution as in the \heavyMax{} model. Nevertheless, we set $\rm\mathcal{A}_{seed}=0.005$, hence obtaining a 4 times smaller MBH number-density than in the \heavyMax{} model.
    The choice of this parameter is  not tied to a particular physical threshold or empirical constraint, but it serves as a controlled variation within the parameter space that provides a more  realistic number density of direct collapse seeds, despite the large uncertainties still surrounding MBH seeding mechanisms and their efficiencies   \citep[see, e.g., the recent work of][]{Cenci2025}. Moreover, it also aligns with our aim to test whether a less efficient seeding scenario remains compatible with both JWST observations and the stochastic GW background detected by PTAs.
  
\end{itemize}

\noindent

Before moving to the core of our analysis, we examine the initial population of MBHs for the four models, given the different assumptions  on seeding.  Fig.~\ref{fig:Newly_Formed_Seed_All_Models} shows the number density of newly formed seeds as a function of redshift. As expected, the \supereddington{} and \eddington{} models predict identical trends, since they employ the same seeding prescription. 
In these two models, light seeds are largely predominant over all other seed types at all redshifts. The number density of newly-formed light seeds increases monotonically down to $z\,{\sim}\,9.5$, where it reaches a maximum of $\rm \sim\!0.1\,Mpc^{-3}$. 
Light seeding ceases at $z\,{\sim}\,6$, when the intergalactic gas supply becomes too metal-rich to support PopIII star formation. In the \supereddington{} and \eddington{} models, intermediate-mass seeds represent the second most common seed type. These start forming at approximately $z\,{\sim}\,8$, when physical conditions in the intergalactic medium become favorable. Intermediate seed formation peaks at $z\,{\sim}\,7.5$, reaching values of about $\rm 10^{-4} \, Mpc^{-3}$. Finally, heavy seeds are the rarest type of MBHs forming in both the \supereddington{} and \eddington{} models:  heavy seed formation begins at $z\,{\sim}\,13$ and it reaches a maximum of about $\rm 10^{-5} \, Mpc^{-3}$ at $z\,{\sim}\,9$. This is in line with the predictions presented in \cite{Dijkstra2014} for the case with moderately high illumination by UV backgrounds (lowest gray, dotted line with stars in all panels of Fig.~\ref{fig:Newly_Formed_Seed_All_Models}). Similarly, this number density also qualitatively agrees with the model variations presented in \cite{Habouzit2016} (gray diamonds in all panels). As detailed in \cite{Spinoso2022}, $z\,{\sim}\,9$ marks the last suitable epoch for the formation of heavy seeds, due to the absence of chemically-pristine regions in our simulated intergalactic medium.

Moving to the  \heavyMax{} and \heavyMin{} models (bottom panels of   Fig.~\ref{fig:Newly_Formed_Seed_All_Models}), as introduced above, these two models only admit the formation of heavy seeds. In both cases, the number density of newly-formed seeds follows the same evolutionary trend, although with different normalizations. This trend initially increases with decreasing redshift, reaching its maximum at $z\,{\sim}\,9.5$, i.e., at a similar moment as for the light seeds in the \supereddington{} and \eddington{} models (see Fig.~\ref{fig:Newly_Formed_Seed_All_Models}). In the \heavyMax{} model, newly-formed heavy seeds reach a maximum number density of ${\sim}\,0.1 \, \rm Mpc^{-3}$, roughly corresponding to a formation rate of $\rm {\sim}\, 2\,{\times}\,10^{-3} Mpc^{-3} Myr^{-1}$. On the other hand, the \heavyMin{} model produces a maximum density of ${\sim}\,0.025\, \rm Mpc^{-3}$, corresponding to a rate of roughly $\rm {\sim}\, 5\,{\times}\,10^{-4} Mpc^{-3} Myr^{-1}$. 
These values are higher by ${>}\,1.5$ and ${>}\,0.5$ dex (for the \heavyMax{} and \heavyMin{} models, respectively) than the ones predicted by the most optimistic scenario presented in \cite{Dijkstra2014}. In detail, this scenario employs a low threshold of UV illumination\footnote{Illumination by photons in the Lyman-Werner UV band is considered necessary for the formation of heavy seeds, since it effectively raises the Jeans mass of primordial, pristine gas clouds, hence hindering gas fragmentation, star formation and the production of light  seeds.} hence strongly favoring the formation of heavy seeds, up to levels generally considered unrealistic \citep[see][for further details]{Dijkstra2014}. This shows that the number density of heavy seeds we obtain for both the \heavyMax{} and \heavyMin{} models can be effectively considered as a \textit{boosted} upper limit to the number density of heavy seeds, at least according to current MBH seeding models.
We stress again that we purposely impose the \heavyMax{} and \heavyMin{} to produce boosted number densities of heavy seeds. Indeed, the main scientific goals of our work are to verify whether recent JWST observations and PTA results can be explained by an MBH population born entirely from heavy seeds. \\

\begin{figure}
    \centering
    \includegraphics[width=1.0\columnwidth]{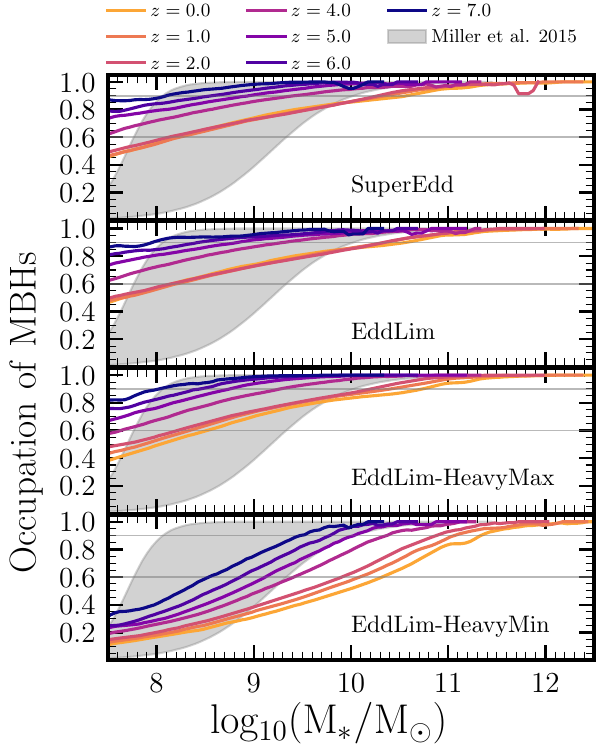}
    \caption{Redshift evolution of the occupation fraction of MBHs in galaxies with different stellar masses. The results correspond to \lgbh{} when run in the \MSII{} merger trees and are compared with the constraints of \cite{Miller2015} (grey shaded area). Horizontal grey lines highlight the 0.5 and 0.9 values.}
    \label{fig:OcupationFraction_All_Models}
\end{figure}

The different seeding mechanisms described above have a significant impact on the occupation fraction of MBHs within galaxy nuclei. This is illustrated in Fig.~\ref{fig:OcupationFraction_All_Models}, which shows the MBH occupation fraction, defined as the proportion of galaxies within a given mass bin that host a central MBH. 
As we can see, the \supereddington{} and \eddington{} models exhibit similar trends, as they include an identical seeding approach. Specifically, at redshifts $z\,{>}\,5$, the occupation fraction exceeds 80\% for any galaxy with a stellar mass $\mstar \,{>}\, 10^8\, \msun$. However, this trend changes at lower redshifts ($z\,{<}\,2$): for systems with $\rm 10^8 \,{<}\,  \mstar \,{<} \, 10^{10} \, \msun$, the occupation fraction varies from 60\% to 80\%. 
Conversely, in more massive galaxies, the occupation fraction is consistently above 80\%, reaching 100\% for galaxies with $\mstar \,{>}\,10^{10.75}\, \msun$.

The \heavyMax{} model exhibits  trends very similar to those seen in the two previous models. This is mainly because the number density of heavy seeds is comparable to that of light seeds in the \supereddington{} and \eddington{} models. Consequently, the occupation fraction is primarily influenced by how frequently seeds form, rather than the mass at which these seeds are born. This becomes even more apparent in the \heavyMin{} model, where lower occupation fractions compared to \heavyMax{} result in significantly different overall occupation fractions. For example, at $z\,{<}\,2$, approximately 60\% of galaxies with stellar masses around $\rm 10^{10} \, \msun$ host a nuclear MBH, raising a tension with the $z\,{\sim}\,0$ constraints of \cite{Miller2015}. It is worth noting that the  our models predict slightly lower occupation fractions at stellar masses around $\mstar \,{\sim}\,10^{10}\,{-}\,10^{10.75}\, \msun$ than those indicated by the constraints from \cite{Miller2015}. This is primarily because gravitational recoils can  displace MBHs from the nuclei of these galaxies, which are mainly characterized by small pseudobulge structures \citep[see a discussion on this in ][]{IzquierdoVillalba2020}. 
 
After exploring the occurrence of MBH formation and the fraction of them that end up in the nuclei of galaxies, in the next section we finally enter in the main part of the paper, examining the predictions of each model regarding the recent discoveries made by JWST and PTA experiments.

\section{Constraining models with JWST and PTA} \label{sec:model_constraints}

This section presents the core results of our work. We begin by demonstrating that JWST observations of the high-$z$ population of active black holes require a substantial number density of  MBHs to already be in place by $z\,{=}\,5$. As we will show, this can be  achieved either via phases of super-Eddington growth or via the formation of a very large population of heavy seeds. Then, we use the recent results on the sGWB from PTAs to further constrain MBHs models and reduce this degeneracy.

\subsection{Constraints from high-z JWST data}

We begin by showing the predictions of our four models for the high-$z$ population of MBHs and compare them with recent estimates from JWST observations. We will show how our physically-motivated seeding model coupled with an Eddington-limited accretion fails at reproducing the large number densities of accreting black holes observed by JWST, while the model that allow for phases of Super-Eddington accretion and the models with a boosted heavy-seeding provide a better description of the data.

\subsubsection{AGN Luminosity Function}\label{sec:AGN_LF_results_analysis}

\begin{figure}
    \centering
    \includegraphics[width=1.0\columnwidth]{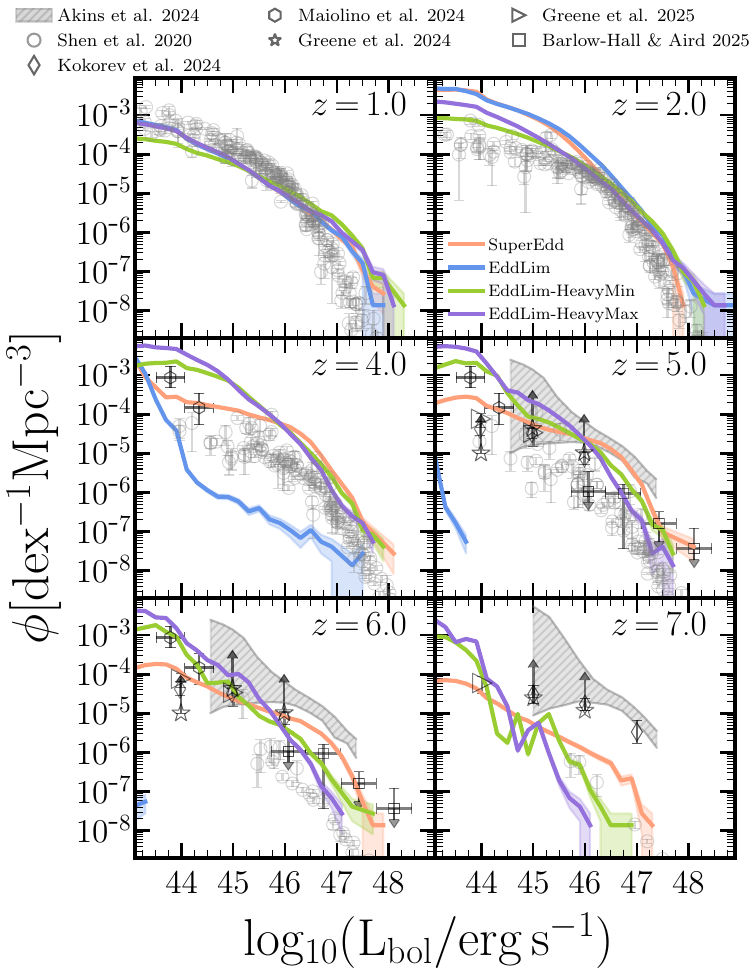}
    \caption{Bolometric AGN luminosity function at $z\,{=}\,1,2,4,5,6,8$ when \lgbh{} is run on the \MSG{}. Orange, blue, green and purple lines correspond to the results of the \supereddington{}, \eddington{}, \heavyMax{} and \heavyMin{} models, respectively. The predictions are compared with the observations of \cite{Shen2020} (compilation of infrared, optical, UV, and X-ray data), \cite{BarlowHall2025} (X-ray data) and \cite{Greene2024,Kokorev2024,Maiolino2024,Akins2024, Greene2025} (JWST data).  We note that the most recent estimates of \citet{Greene2025} (triangles), derived using a new bolometric correction, bring the data more consistent with our predictions, especially at $z=7$. }
    \label{fig:QLF_All_Models}
\end{figure}

We start by looking at our predictions for the bolometric luminosity function and compare them with pre-JWST AGN/QSO data and recent JWST observations.  Indeed, current observations from JWST are providing completely new insights into the luminosity distribution of active MBHs at $z\, {>} \,5$, complementing the quasar luminosity function previously estimated for only the brightest AGN. In Fig.~\ref{fig:QLF_All_Models} we show the predictions of our four models  across redshift and for a broad range of luminosities. We compare our predictions with the bolometric quasar luminosity function derived by  \cite{Shen2020} using a compilation of infrared, optical, UV, and X-ray data. For $z \,{\sim}\,5\,{-}\,7$ we also add the bolometric luminosity function recently estimated by \cite{BarlowHall2025} for intermediate-luminosity AGN with Chandra COSMOS data.  JWST results from \cite{Maiolino2024}, \cite{Akins2024}, \cite{Kokorev2024}, \cite{Greene2024} and \cite{Greene2005} are shown for $z\,{\gtrsim}\, 4$. These include both galaxies selected for their broad-lines indicative of AGN emission \citep[][]{Greene2024, Maiolino2024, Greene2025} as well as LRDs \citep[][]{Kokorev2024, Akins2024}, assuming that they are all AGN.

  At $z\, {=} \, 1$ and $z\, {=} \, 2$ all models give similar predictions for the luminosity function and are all similarly consistent with observations. It is at higher redshifts where the models start differing substantially with each other. First of all, the \eddington{} model significantly underestimates the number density of AGN over the entire luminosity range, with effectively no objects with $\rm L_{\rm Bol} \, {>} 10^{43} \, ergs/s$ at  $z\, {>} \, 5 $. The other three models, instead, are fairly consistent with the data and, in particular, they all predict a substantial population of active MBHs also at the faint end, consistent with the recent JWST observations. We also note some differences between these three models: the \heavyMax{} and \heavyMin{} models predict a steeper luminosity function, with very large number densities for  $\rm \mathit{L}_{bol}\,{<}\, 10^{46}\, erg/s$ and a drop at larger luminosities. The \supereddington{} model, instead, predicts a shallower dependence with luminosity, leading to a larger number of bright objects at $z=7$.  While our models predict an important redshift evolution of the number density of accreting MBHs, consistent with previous quasars observations, the recent JWST results seem instead to point to a more constant luminosity function, at least until $z  \, \sim \, 7$. We note, however, that the estimates of \cite{Akins2024} have been derived assuming that all LRDs in their sample are AGN, and the ones of  \cite{Greene2024} are based on $3$ objects at $z\, \gtrsim \, 7$.  The more recent estimates of \citet{Greene2025}, derived using a new bolometric correction, bring the data leftwards, making it more consistent with our predictions (see the horizontal triangles in the figure). Cleaner and more complete samples will be needed to confirm these trends.

\subsubsection{Black Hole Mass Function}\label{sec:BHMF_results_analysis}
\noindent

\begin{figure}
    \centering
    \includegraphics[width=1.0\columnwidth]{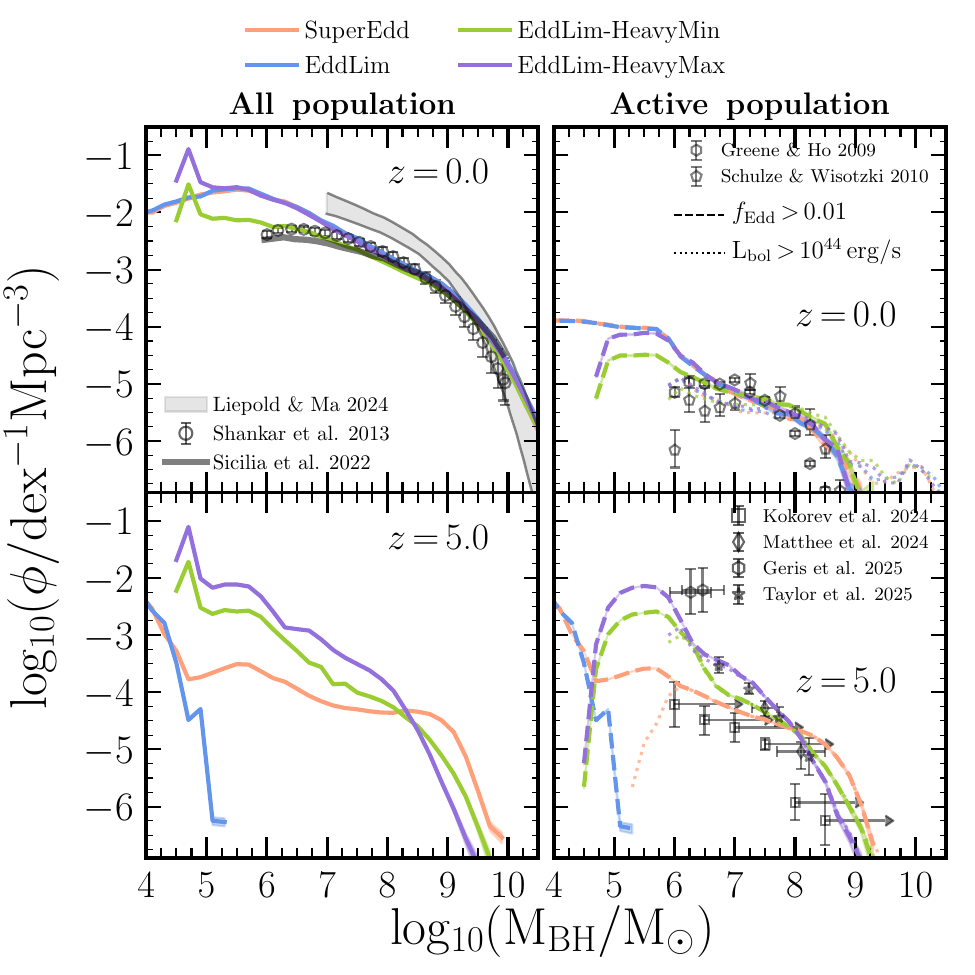}
    \caption{Black hole mass function at $z\,{=}\,0$ (top panels) and $5$ (bottom panels), predicted by the \supereddington{}, \eddington{}, \heavyMax{} and \heavyMin{} models. The left panels show the mass functions for the full black hole population and are compared with the constraints of \cite{Shankar2013,LiepoldAndMa2024} and \cite{Sicilia2022}. The right panels show only the mass function for active massive black holes, defined as the MBHs with either an Eddington ratio $f_{\rm Edd}\,{>}\, 0.01$ (solid lines), or with a bolometric luminosity $\rm \mathit{L}_{\rm Bol}\,{>}\,10^{44} erg/s$ (dashed lines). The results are compared with \cite{Greene2007,Schulze2010} ($z\,{=}\,0$),  \cite{Matthee2024,Kokorev2024,Geris2025} (JWST data, $z\,{=}\,5$) and \cite{Taylor2025} ($z\,{\sim}\,4$). }
    \label{fig:BHMF_All_Models}
\end{figure}

We now move to the black hole mass function (BHMF) predicted by our four models. We show it in Fig.~\ref{fig:BHMF_All_Models}, both at  $z\,{=}\,0$ and $z\,{=}\,5$. To fairly compare the model results with observational constraints, we show the mass function of the entire MBH population (left column) and the one of  \textit{active} MBHs only  (right column).\\

Starting with the predictions for the full $z\,{=}\,0$ MBH population (top left panel)
 of  Fig.~\ref{fig:BHMF_All_Models}, we note that all the models closely align with the observational constraints of \cite{Shankar2013} as well as the more recent estimates of \citet{Sicilia2022} (derived from a semi-empirical method that effectively uses the continuity equation) and of \citet{LiepoldAndMa2024} (who derived the black hole mass function from the galaxy stellar mass function via scaling relations). All models predict a slightly larger number density for $\rm \mathit{M}_{BH}\,{>}\,10^{9}\,\msun$ with respect to observations. As we will discuss later, a larger  number density of MBHs at these high masses is needed to predict a sGWB consistent with current PTA data (see Sect.~\ref{sec:PTA}). 
 In the mass range $\rm 10^{\,6.5\,}\msun\,{<}\, \mathit{M}_{BH}\,{<}\,10^{\,8.5\,}\msun$ all models are more consistent with the estimates of \cite{Shankar2013}, while the BHMF of \cite{LiepoldAndMa2024} has a larger normalization in this mass range. Note that the \heavyMin{} model is the one predicting a lower BHMF in the range $\rm \mathit{M}_{BH}\,{<}\, 10^{7}\,\msun$ because of the lower occupation fraction in intermediate and low-mass galaxies (see Fig. \ref{fig:OcupationFraction_All_Models}). The peaks that the models \heavyMax{} and \heavyMin{} show for $\rm M_{BH}\,{\sim}\,10^{5}\,\msun$ are due to massive seeds that have not been able to grow since formation.
 
 The picture presented above changes substantially when we move to the predictions for $z\,{=}\,5$ (bottom left panel of Fig.~\ref{fig:BHMF_All_Models}), revealing the profound impact of MBH formation and evolution processes on the statistical properties of the high-$z$ MBH population. In particular, \supereddington{} is the only model variation able to produce a significant population of MBHs with $\rm M_{BH}\,{>}\,10^9\, \msun$, in line with observations of high-$z$, luminous QSOs \citep[e.g.][]{farina2022}. On the contrary, the \eddington{} variation struggles to reach even $\rm M_{BH}\,{\sim} \,10^6\, \msun$. When considering this result together with the analysis shown in Fig.~\ref{fig:Newly_Formed_Seed_All_Models}, it is clear that a MBH population formed predominantly as light BH-seeds cannot reach the masses required to explain the high-$z$ AGN  population by exclusively accreting gas in the Eddington-limited regime. However, as expected, this constraint is easily evaded if MBHs form as heavy seeds only. Indeed, both the \heavyMax{} and \heavyMin{} model variations manage to reach $\rm \mathit{M}_{BH}\, {>}\, 10^8\, \msun$ even with  Eddington-limited accretion.\\

Moving to the mass function of active black holes (panels in the right column of Fig.~\ref{fig:BHMF_All_Models}), we   show our predictions using two simple definitions for the population of active MBHs: massive black holes are assumed to be active either if (i) $f_{\rm Edd}\,{>}\,0.01$ or if  (ii) $\rm \mathit{L}_{bol} \,{>}\, 10^{44} \, erg/s$ (this  luminosity threshold corresponds to the typical lower-limit of AGN luminosities observed by JWST). On the observational side, the black hole masses powering AGN can be estimated using reverberation mapping  \citep[e.g.,][]{Peterson2004} or  single-epoch virial relations \citep[e.g.,][]{Kaspi2000, Greene2005, Vestergaard2006, Trakhtenbrot2012}.   Starting with the active population at $z\,{=}\,0$, the four model variations are compatible with each other and show a good agreement with the local constraints of \cite{Greene2007} and \cite{Schulze2010}. In particular, all models are able to capture the flattening of the active BHMF for $\rm M_{BH}\leq10^{7\,}\msun$ as well as its \textit{knee} at $\rm M_{BH}\sim10^{8\,}\msun$, independently on the definition used for active black holes. Note that when we use the cut based on bolometric luminosity (dotted lines), the population of active MBHs reaches higher masses, as very massive MBHs can shine at  high luminosities even when accreting at relatively low Eddington rates.

Despite being calibrated at lower redshifts, the phenomenological virial relations mentioned above have also been used on JWST spectra to estimate the masses of the MBHs powering high-z active galaxies. The estimates of  \cite{Matthee2024} are shown in the bottom-right panel of Fig.~\ref{fig:BHMF_All_Models}, together with the lower limits of \cite{Kokorev2024} who, instead, derive the active black hole mass function from the bolometric luminosity function assuming that the black holes are accreting at the Eddington rates. 
 Comparing these estimates  with our model predictions,  we can see that the \supereddington{}, \heavyMax{} and \heavyMin{} models agree overall well with JWST data. Similarly as for the case of the LF, we see that the \supereddington{} model is the one producing the largest population of active black holes with $\rm \mathit{M}_{BH}\, {>}\, 10^8\, \msun$. 
 This is an indication that the \supereddington{} model tends to favor an earlier assembly of MBHs, compared with all the other models. The \heavyMax{} and \heavyMin{} models, however, predict a larger number density of more moderate mass black holes ($\rm 10^5\,{<}\,  \mathit{M}_{BH}\, {<}\, 10^7\, \msun$). Larger number densities  in this mass ranges have also been recently estimated by \cite{Geris2025} by using stacked spectra of low-luminosity AGN from the JADES survey, possibly compensating the incompleteness at the low-mass end from previous studies \citep[see also the discussion in ][]{Matthee2024}.

In summary, the BHMF predictions of the \supereddington{}, \heavyMax{} and \heavyMin{} models are all compatible with current observational constraints of the BHMF derived in the local and high-$z$ Universe for the full population and the active only. This, together with the fact that the \eddington{} model cannot produce a population of MBHs with masses $\rm \mathit{M}_{BH}\,{>}\,10^6\,\msun$ at high-$z$ (either active or inactive) enables us to disfavor the \eddington{} model: if MBHs are only allowed to grow via Eddington-limited accretion, light seeds cannot reach $\rm \mathit{M}_{BH}\,{>}\,10^6\,\msun$ at $z\,{\geq}\,5$, while RSM and DCBH seeds are simply too rare to play a significant role in the build-up of the massive-end of the BHMF.  We stress that the \supereddington{}, \heavyMax{} and \heavyMin{} models give all similar estimates for the BHMF at the high-mass end. Clearly, it is in the low-mass regime where degeneracies can be broken, and more data will be needed to constrain our theoretical assumptions.

\subsubsection{Scaling relations}\label{sec:ScalingRelation_results_analysis}

\begin{figure}
    \centering
    \includegraphics[width=1.\columnwidth]{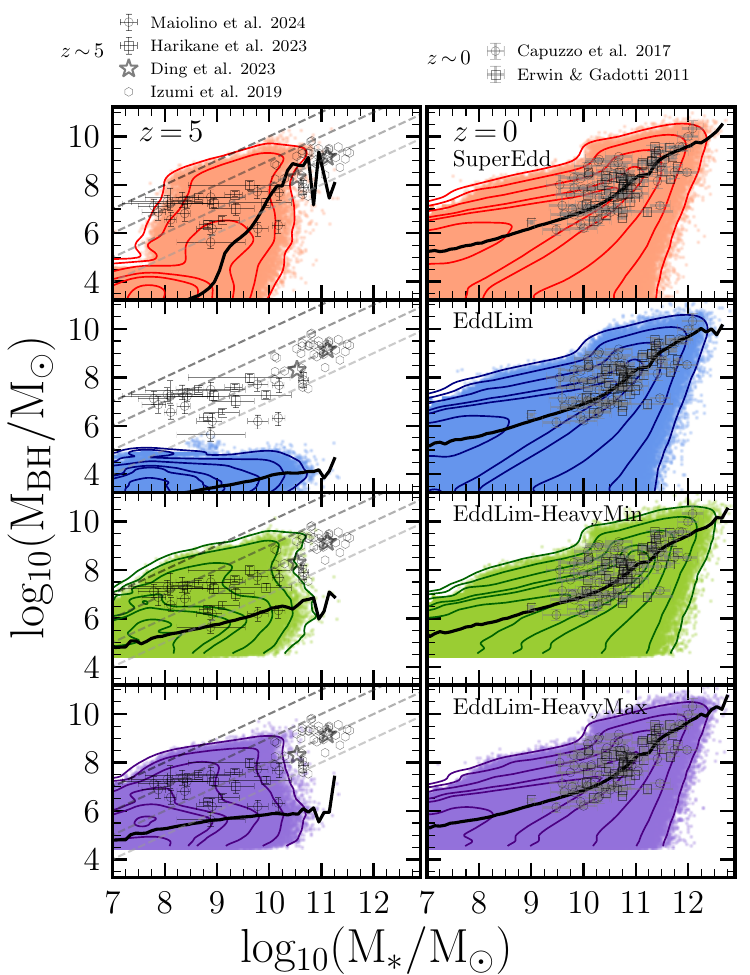}
    \caption{$\rm \mathit{M}_{BH}\,{-}\, \mstar$ relation  at $z\,{=}\,0$ (right) and $5$ (left) when \lgbh{} is run on the \MSG{}. Orange, blue, purple and grey colors correspond to the results of the \supereddington{}, \eddington{}, \heavyMax{} and \heavyMin{} models, respectively. 
    Colored points show the outputs of the runs, colored contours represent the density distribution of the points, and black lines represent the medians. Grey dashed curves guide the eyes from a $1\,{:}\,1$ relation to a $1\,{:}\,10^{3}$ scaling. Grey symbols indicate observational estimates of \cite{Maiolino2024,Harikane2023,Ding2023,Izumi2019} ($z\,{=}\,5$) and \cite{Capuzzo2017,Erwin2012} ($z\,{=}\,0$).}
    \label{fig:ScalingRelations_All_Models}
\end{figure}

We now move to our predictions for the scaling relation between black hole and stellar mass. The high black hole to stellar mass ratios inferred at very high-$z$ have indeed been one of the most surprising results of the first JWST observations. In Fig.~\ref{fig:ScalingRelations_All_Models} we show our model predictions at $z=5$,  together with the results of \cite{Harikane2023, Ding2023} and \cite{Maiolino2024}. We also add the estimates from \cite{Izumi2019}, based on the low-luminosity quasar sample derived from the SHELLQs survey \citep{Matsuoka2016}.  The \supereddington{}, \heavyMax{} and \heavyMin{} models all predict a quite large scatter in the scaling relation, and are able to produce the high-$z$ massive broaed-line systems seen by JWST and a few objects reaching the large masses of quasar surveys.  However, when we compare the median values of the $\rm \mathit{M}_{BH}\,{-}\, \mstar$ relation (black lines), the three models predict different trends: the \supereddington{} model shows a steep relation  at $\mstar\,{>}\,10^9\, \msun$, while for the \heavyMax{} and \heavyMin{} models the median of the relation remains quite flat. Indeed, as we have previously seen for the BHMF at this redshift, the  \heavyMax{} and \heavyMin{} models predict a larger number of low-mass MBHs with respect to the  \supereddington{} model.  We underline that here we are including all MBHs. A more careful comparison with JWST results, including only our active black holes, will be discussed in Section~\ref{sec:prefModels}. We will also discuss the possibility of a tighter relation with the dynamical mass of the galaxy.

In the right panel of Fig.~\ref{fig:ScalingRelations_All_Models}, we show the scaling relations at $z\,{=}\,0$. As for the black hole luminosity and mass functions,  all the models (including the \eddington{} one) predict similar trends, that are also  in agreement with the observational estimates from \cite{Capuzzo2017} and \cite{Erwin2012}. Confirming what we have seen for the AGN luminosity and black hole mass functions,  all models give similar predictions for the low-redshift population, and are all consistent with observational estimates. It is in the high-$z$ Universe where differences appear, and where degeneracies between models can be broken. \\

Based on the results presented in this section, along with the BHMF and LFs discussed above, we can robustly conclude that the most recent JWST observations rule out the \eddington{} model. Phases of super-Eddington accretion or a boosted and early rising population of heavy seed MBHs  are needed to explain the large number densities of massive and luminous black holes derived from JWST surveys.

\subsection{Constraints from PTA} \label{sec:PTA}

As shown in the previous section, the electromagnetic constraints from JWST enable us to rule out the \eddington{} model. However, the results for the \supereddington{}, \heavyMax{}, and \heavyMin{} models remain consistent with the observed abundance and properties of high-$z$ and low-$z$ MBH populations. To better distinguish between these remaining models, we will now examine their GW signals in the nano-Hz frequency range. This new observational window has recently been opened up by PTA collaborations, offering a novel way to study the assembly and evolution of MBHs. In particular, the European Pulsar Timing Array \citep[EPTA,][]{Desvignes2016}, the North American Nanohertz Observatory for GWs \citep[NANOGrav,][]{Arzoumanian2015}, the \textit{Parkes} Pulsar Timing Array \citep[PPTA,][]{Reardon2016} and the Chinese Pulsar Timing Array \citep[CPTA,][]{Lee2016} collaborations have reported evidence at a significant level of $2\,{-}\,4\sigma$ for a nano-Hz sGWB compatible with the existence of low-$z$ MBHBs with masses $\,{>}\,10^{8}\, \msun$ \citep{Afzal2023,InterpretationPaperEPTA2024}. Assuming circular binaries, the reported sGWB has an amplitude that ranges between $[1.7\,{-}\,3.2] \,{\times}\,10^{-15}$, at a reference frequency $f\,{=}\,1 \, \rm yr^{-1}$ \citep[][]{Antoniadis2023,Agazie2023,Reardon2023,Xu2023}. To compare our predictions with these PTA constraints, we follow \cite{Sesana2008} expressing the characteristic sGWB from a population of inspiralling MBHBs as:
\begin{equation} \label{eq:GWB1}
    h^2_c(f) \, {=} \, \frac{4 G^{5/3}}{f^{2} c^2\pi} {\int}{\int} \frac{dz d\mathcal{M}}{(1+z)} \frac{d^2n}{dzd\mathcal{M}}\frac{d\mathrm{E}_{\rm GW}(\mathcal{M})}{d\ln{f_r}},
\end{equation}
where $d^2n/dzd\mathcal{M}$ is the comoving number density of MBHB mergers  per unit redshift, $z$, and  $\mathcal{M}$ is the rest-frame chirp mass\footnote{ The chirp mass of a MBHB system is expressed as ${\mathcal{M}} \,{=}\, \rm (M_{\rm BH,1}M_{\rm BH,2})^{3/5}/(M_{\rm BH,1}+M_{BH,2})^{1/5}$ where $\rm M_{\rm BH,1}$ and $\rm M_{\rm BH,2}$ are the masses of the primary and secondary MBH, respectively.}, and $f$ is the frequency of the GWs in the observer frame. The quantity $d\mathrm{E_{GW}}/d\ln{f_r}$ represents the differential energy spectrum of the binary, i.e the energy emitted per logarithmic rest-frame frequency, $f_r$. With the assumptions that the MBHBs in the PTA band are purely driven by GW emission (i.e., no coupling with the environment) and in perfectly circular orbits, Eq.~\eqref{eq:GWB1} can be expressed as:
\begin{equation} \label{eq:SGWB}
    h^2_c(f) \, {=} \, \frac{4 G^{5/3} f^{-4/3}}{3c^2\pi^{1/3}} {\int}{\int}  \frac{d^2n}{dzd\mathcal{M}} \frac{\mathcal{M}^{5/3}}{(1+z)^{1/3}} \, dz \, d\mathcal{M} ,
\end{equation}
often re-written as:
\begin{equation} \label{eq:SGWB2}
h_c(f) \,{=}\, A\left( \frac{f}{f_0} \right)^{-2/3} ,
\end{equation}
where $A$ is the amplitude of the signal at the reference frequency $f_0$ set to $\rm 1yr^{-1}$. Therefore, by taking $d^2n/dzd\mathcal{M}$ and using Eq.~\eqref{eq:SGWB} and Eq.~\eqref{eq:SGWB2} we can determine the sGWB predicted by our  models.\\

\begin{figure}
    \centering
    \includegraphics[width=1.0\columnwidth]{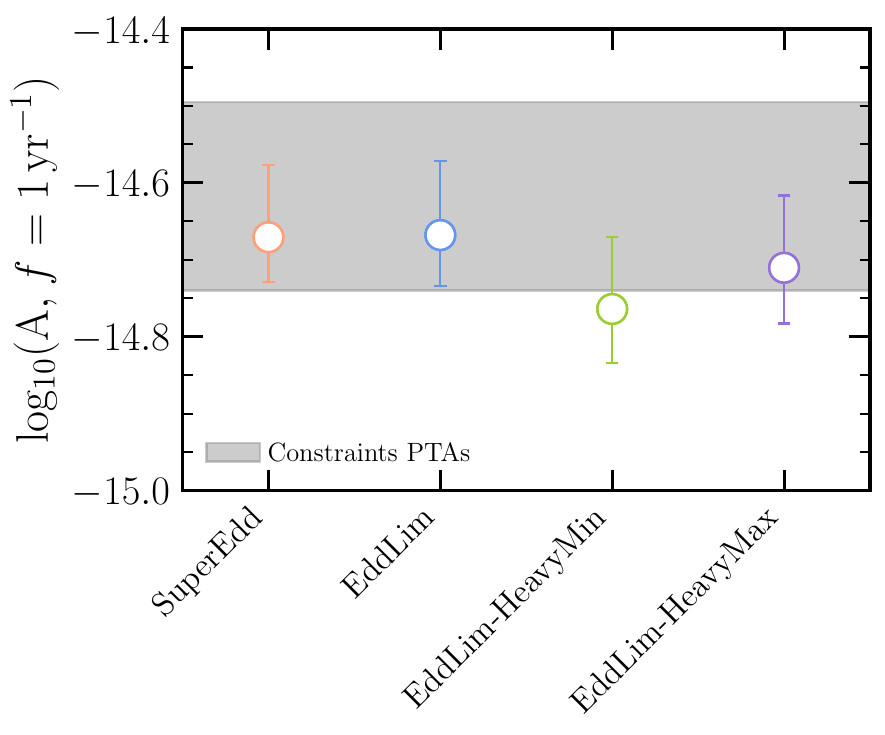}
    \caption{Stochastic GWB amplitude at $\rm 1\, yr^{-1}$ reference frequency predicted by the four different models when \lgbh{} is \MSG{}. The errors on the sGWB amplitude have been computed by dividing the MS box into sub-boxes of $\rm 100 \, Mpc$ side-length. Then, it was computed with all of them the $\rm 16^{th}$ and $\rm 84^{th}$ percentile of $\rm A\,@1\,yr^{-1}$. The shaded area represents the observational constraints reported by EPTA, NANOGrav, PPTA and CPTA collaborations \citep[][]{Antoniadis2023,Agazie2023,Reardon2023,Xu2023}.}
    \label{fig:sGWB}
\end{figure}

In Fig.~\ref{fig:sGWB} we show the amplitude that we derive for the population of MBHBs predicted by our four models, and compare it with the constraints of PTA experiments. The \supereddington{} and \eddington{} models (orange and cyan points) are the ones that align the best with the observed PTA constraints. The \heavyMax{} model (purple point) is also compatible with PTA results but predicts an amplitude that populates the lower end of the constraints. The \heavyMin{} model (green point) is the only one that falls outside the estimated range for the sGWB, indicating that it is less compatible with current data. This tension suggests that a sustained formation of heavy seeds of $\rm {\sim}\, 0.01 \, Mpc^{-3}$ between $7\,{<}\,z\,{<}\,14$ (i.e $\rm {\sim}\,5\,{\times}\,10^{-4}\, Mpc^{-3}\, Myr^{-1}$ formation rate) predicted by \heavyMin{} is not sufficient to produce a population of low-$z$ MBH and MBHB compatible with PTA observations. These results highlight the significant potential of current and future GW experiments to constrain models of MBH formation and growth. This was already discussed in \cite{IzquierdoVillalba2021} using a previous version of our \lgbh{} model. The authors concluded that to obtain a large enough amplitude of the sGWB, a large population of SMBHs needs to be assembled \citep[see also ][]{SatoPolito2023}. The recent estimates of the massive end of the BHMF are encouraging in this direction \citep[see ][]{LiepoldAndMa2024}.\\

An important caveat to consider in our analysis is the assumption that the  MBHBs that give rise to the nHz sGWB, evolve solely due to GWs. The recent PTA results suggest that the observed spectral slope of the signal is flatter than the standard value of $-2/3$ expected for a population of circular MBHBs driven only by GW emission (see Eq.~\ref{eq:SGWB2}). Although significant uncertainties are preventing definitive conclusions, this deviation could have important implications for the properties of the underlying MBHB population. Specifically, it may indicate a strong coupling with their stellar environment during their inspiralling phase \citep{Quinlan1997, Sesana2006}. Alternatively, it may suggest that binaries in the PTA band possess non-negligible orbital eccentricities \citep[e.g.,][]{Gualandris2022, Fastidio2024}. In addition to the spectral slope, the recent PTA signal appears to be louder than what current cosmological hydrodynamical simulations and semi-analytical models predict \citep[see e.g.,][]{Sesana2009,Kelley2016,Kelley2017,Bonetti2018,IzquierdoVillalba2021,Curylo2023,Saeedzadeh2023,Li2024,Chen2025}. Several approaches have been proposed to explain these high signal levels. For example, large sGWB amplitudes can be achieved by assuming a rapid dynamical evolution of MBHBs, with timescales of ${<}\,$1 Gyr. Alternatively, a faster and more substantial mass growth of MBHs or a higher normalization in the scaling relations can also produce stronger signals \citep[see e.g.,][]{InterpretationPaperEPTA2023,Agazie2023}. Noticeably, the latter statement is closely related to the relatively high sGWB predicted in \lgbh{}, regardless of the particular model: \lgbh{} generates at $z\,{=}\,0$ a number density of MBHs with masses $\rm {>}\, 10^8\, \msun$ relatively larger with respect to current observational estimates (see Fig.~\ref{fig:BHMF_All_Models}).\\

In summary, the current PTA estimates of the sGWB enable us to exclude the \heavyMin{} model and establish a lower limit on the number density of heavy seeds.

\section{MBH properties of preferred models/predictions} \label{sec:prefModels}

Based on the analysis presented in the previous section, a substantial population of MBHs with $\rm M_{BH} \,{>}\,10^6 \, \msun$ must have been in place already at very high redshifts to  explain JWST observations (see Fig.~\ref{fig:QLF_All_Models}, Fig.~\ref{fig:BHMF_All_Models} and Fig.~\ref{fig:ScalingRelations_All_Models}). Additionally, the nano-Hertz sGWB reported by PTAs provides further constraints on our models, requiring a high occupation fraction of MBHs in galaxies with stellar masses $M_{*}\,{>}\,10^{10}\, \msun$ at low redshifts (see Fig.~\ref{fig:OcupationFraction_All_Models} and Fig.~\ref{fig:sGWB}). Among the four models explored in this work, only the \supereddington{} and \heavyMax{} models meet these requirements. In the \supereddington{} model, MBHs primarily originate from light seeds that undergo episodic super-Eddington accretion, enabling the rapid build-up of a massive population already at high redshift. Conversely, the \heavyMax{} model assumes a highly efficient formation of heavy seeds, resulting in a number density comparable to that of light seeds. In this section, we further investigate the properties of the two preferred models, refine our comparison with JWST observations, and examine the characteristics of the merging MBHB population.

\begin{figure}
    \centering
    \includegraphics[width=1.0\columnwidth]{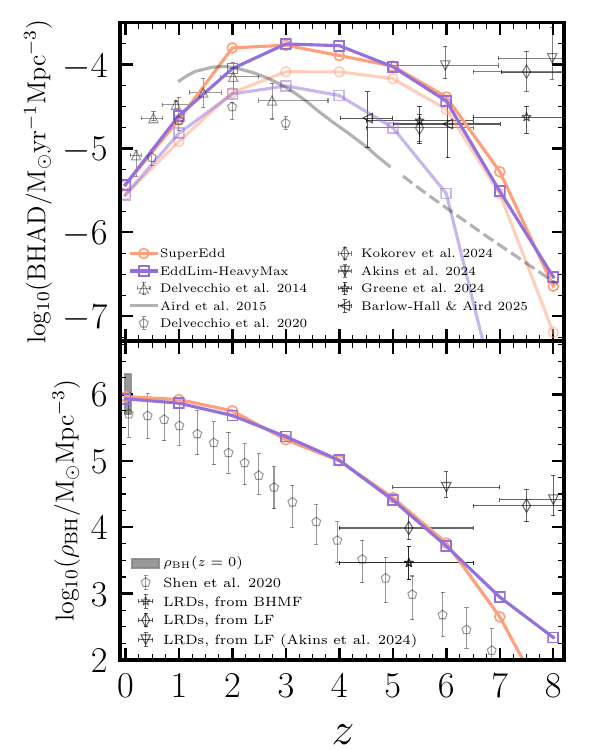}
    \caption{\textbf{Top panel}: Redshift evolution of the black hole accretion rate density predicted by \lgbh{} when run on the \MSG{}. Orange and purple lines correspond to the predictions of the \supereddington{} and \heavyMax{} models, respectively. The pale color lines correspond to the same, but imposing a cut in $\rm \mathit{M}_{BH}\,{>}\,10^8\, \msun$ The results are compared with the observations of \cite{Delvecchio2014,Aird2015,Delvecchio2020,Kokorev2024,Akins2024,Greene2024} and \cite{BarlowHall2025}. \textbf{Bottom panel}: Black hole mass density for the \supereddington{} and \heavyMax{} models. Observational data points correspond to \cite{Shen2020} and the compilation presented in \cite{Inayoshi2024spin}. The current range of estimates at $z\,{=}\,0$ are indicated with the gray boxes \citep[from the compilation of][]{LiepoldAndMa2024}.}
    \label{fig:BHAR_BHMD}
\end{figure}

\subsection{The high-\textit{z} MBH population in the \supereddington{} and \heavyMax{} models}  \label{sec:preferredJWST}

\begin{figure}
    \centering
    \includegraphics[width=1.0\columnwidth]{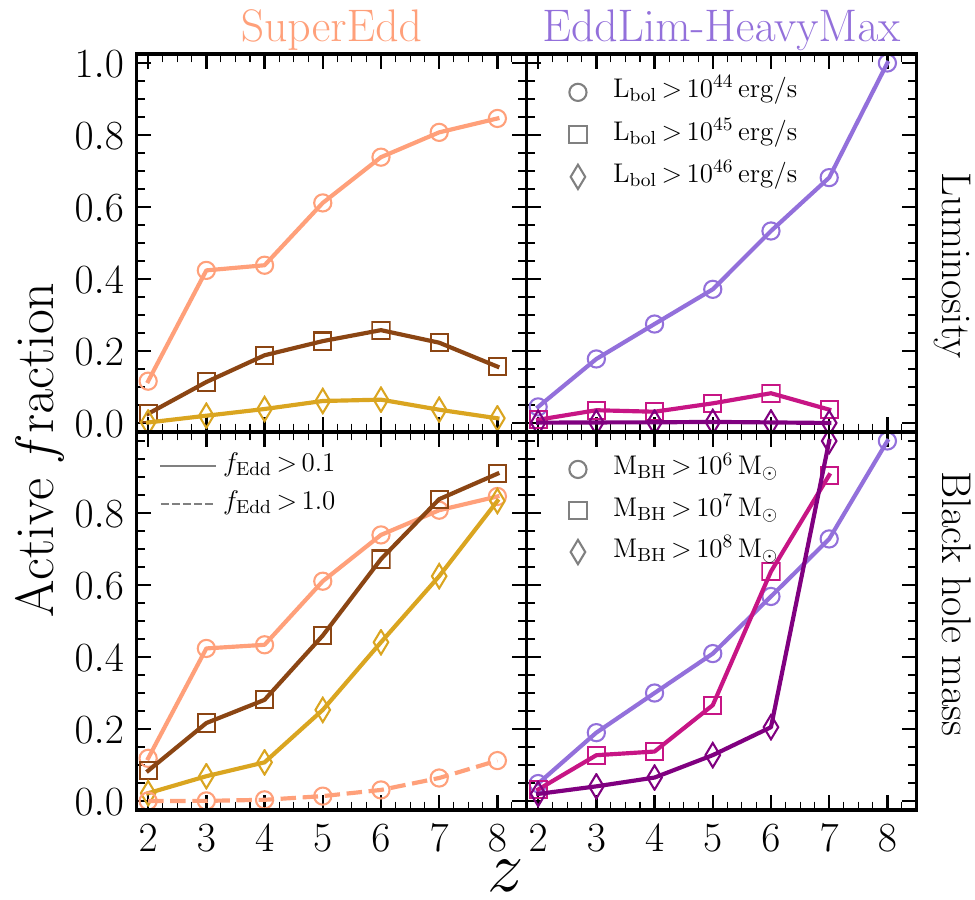}
    \caption{Fraction of active MBHs with $\rm M_{BH}\,{>}\,10^{6}\, \msun$ for the \supereddington{} (left panels) and \heavyMax{} (right panels) models. The upper panel show the results at different luminosity cuts, while the lower ones correspond to different MBH mass cuts. We stress that in the low right panel, no dashed line is included because no super-Eddington is allowed for the \heavyMax{} model.} 
    \label{fig:MBHs_active_fraction}
\end{figure}

\begin{figure*}
    \centering
    \includegraphics[width=0.66\columnwidth]{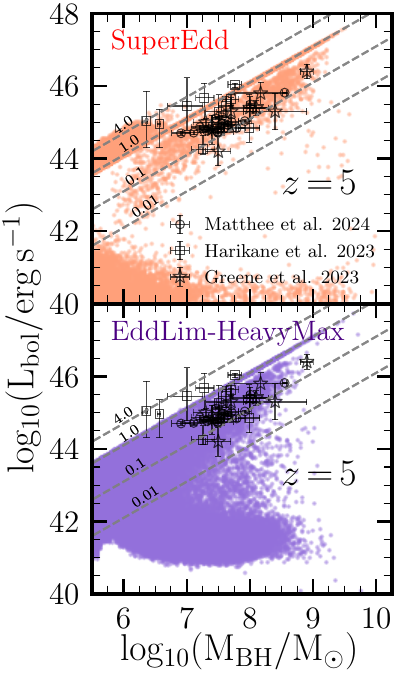}
    \includegraphics[width=1.33\columnwidth]{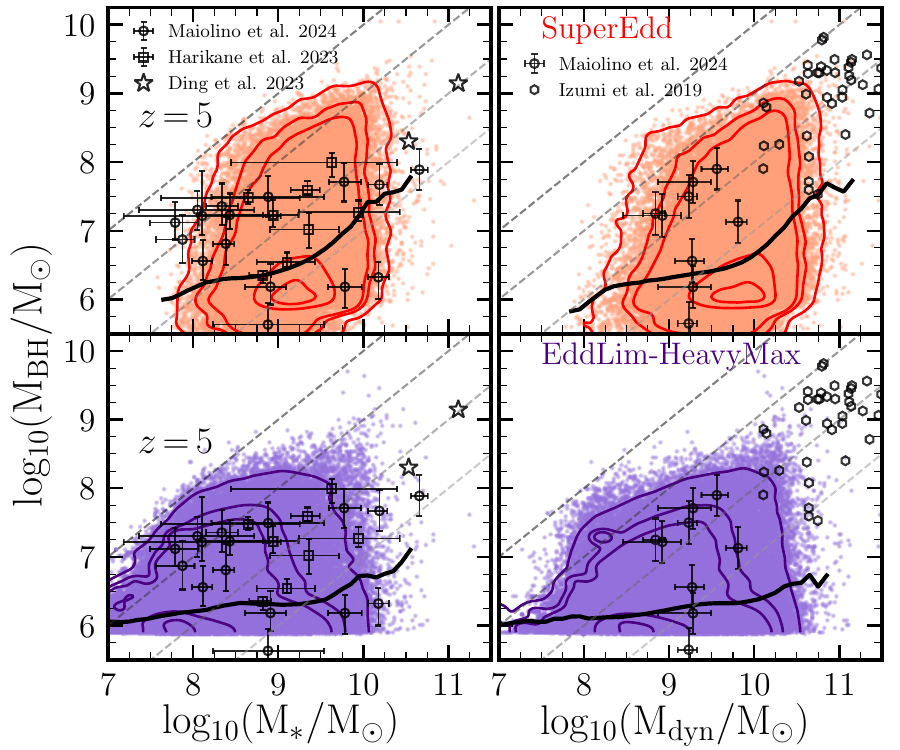}
    \caption{Scaling relations at $z\,{=}\,5$ for the \supereddington{} (top) and \heavyMax{} (bottom) models. \textbf{Left panel}: $\rm L_{bol}\,{-}\,M_{BH}$ plane. The dashed lines correspond to $f_{\rm Edd}\,{=}\,0.01, 0.1, 1, 2, 4$.  The results are compared with the JWST observations presented in \cite{Harikane2023,Matthee2024} and \cite{Greene2024}. \textbf{Middle panel}: $\rm M_{Stellar}\,{-}\,M_{BH}$ plane. The dashed lines correspond to the $1\,{:}\,1$, $1\,{:}\,10$, $1\,{:}\,10^2$ and $1\,{:}\,10^3$ relation.  Colored contours represent the
    density distribution of the points, and black lines represent the medians.The results are compared with \cite{Harikane2023,Ding2023,Maiolino2024} \textbf{Right panel}: $\rm M_{dyn}\,{-}\,M_{BH}$ plane, where $\rm M_{dyn}$ corresponds to the dynamical mass of the galaxy computed as the sum of the stellar and cold gas component. The results are compared with the estimations of \cite{Izumi2019} and {Maiolino2024}.   } 
    \label{fig:scaling_relations_SE_and_HeavyMax}
\end{figure*}

In this section, we study in more detail the predictions of the \supereddington{} and \heavyMax{} models regarding the assembly of MBHs. We start by looking at the redshift evolution of the black hole accretion rate density (BHAD) and the evolution of the resulting black hole mass density ($\rho_{\rm BH}$). These are shown in  Fig.~\ref{fig:BHAR_BHMD}. Again, we compare these with previous data derived for luminous quasars/AGN in various wavelengths \citep{Aird2015, Delvecchio2014, Delvecchio2020, Shen2020, BarlowHall2025}, as well as the recent estimates from JWST broadline galaxies and LRDs \citep[][]{Akins2024, Greene2024, Kokorev2024}. We note that the values for the BHAD are derived by integrating the luminosity function, and the values for the $\rho_{\rm BH}$ for the LRDs are the median ones calculated by \cite{Inayoshi2024spin} either from the BHMF or the LF. 
Focusing first on the BHAD, we do not see significant differences between the two models, except that the \supereddington{}  is peaking at slightly lower redshift than the \heavyMax{}  ($z\, {\sim} \, 2$ instead of $z \, {\sim} \, 4$). However, we note that this is the total BHAD, with no cuts in black hole mass. If we calculate the BHAD only for the most massive black holes ($\rm {>}10^8\,\msun$), the \supereddington{} model shows an early boost in the growth, as indicated by the pale curves in the figure (see also the evolution of the luminosity function shown in Fig. \ref{fig:QLF_All_Models}). When comparing our predictions with observations, it is clear that our models are overall in good agreement with JWST data and the recent deep X-ray data of \cite{BarlowHall2025}. Our models, as well as these new data that include fainter AGN, point to a significantly flatter evolution of the BHAD at high redshift: a major jump in the BHAD happens around $z \, {\sim} \, 6\,{-}\,8$, after which growth is significantly sustained until the drop at $z \, {<} \, 2$.

Regarding the evolution of the black hole mass density (bottom panel of Fig.~\ref{fig:BHAR_BHMD}), we see that both models predict a similar evolution, with significant differences appearing only at $z \, {>} \, 6$. As shown before, for the \supereddington{} model, a large fraction of black holes experience significant growth around this redshift, reaching high masses in a short window of time. As for the BHAD, some of the recent estimates from JWST data suggest a flatter evolution of $\rho_{\rm BH}$ at high redshift, although still with large uncertainties. None of our models show such a flattening as they feature a monotonic decrease towards higher redshifts, but still with a larger normalization with respect to the  estimates derived from the luminosity function of quasars
\citep[e.g.,][]{Shen2020}. In the local  Universe, the values derived from a variety of approaches range from $\rho_{\rm BH}\,  {\sim}  \, 0.5 \,{\times}\, 10^6 \, \rm{M}_{\odot}\, Mpc^{-3}$ to $\rho_{\rm BH}\,  \,{\sim}\,  \, 2 \,{\times}\, 10^6 \, \rm{M}_{\odot} \, Mpc^{-3}$, with the higher values being the most recent estimates from scaling relations  \citep[see the discussion of ][]{LiepoldAndMa2024}. These estimates are broadly consistent with our model predictions, which converge to $\rho_{\rm BH} \,{\sim}\, 10^6 \, \rm{M}_{\odot}\,\mathrm{Mpc}^{-3}$ at $z \,{=}\, 0$.\\

Linked to the black hole mass and accretion rate density is the active fraction, i.e. the fraction of black holes active at a given time. This is connected to the black hole duty cycle, which provides important information on the black hole-galaxy co-evolution \citep[e.g., ][]{Martini2001}. Our predictions for the MBH active fraction are shown in Fig.~\ref{fig:MBHs_active_fraction}. In the top panel, we show the active fraction for MBHs reaching different values of bolometric luminosities. As expected, the fraction of active MBHs depends on luminosity and redshift, with the most luminous ones ($\mathit{L}_{\rm bol} \, {>} \, 10^{46} \, \rm{erg/s}$) having a low active fraction and a mild redshift evolution for both the \supereddington{} and the \heavyMax{} models.
When considering milder luminosities ($\mathit{L}_{\rm bol} \, {>} \, 10^{44} \, \rm{erg/s}$, which are approximately the luminosities of JWST AGN, as discussed before), we see a strong redshift evolution of the active fraction, with most MBHs (active ${\sim} \, 80 \% $) shining above this luminosity cut at $z \, {\sim} \, 6\,{-}\,8$ for the \supereddington{} model.  The active fraction for the \heavyMax{} model is overall lower, with no MBHs with  $\mathit{L}_{\rm bol} \, {>} \, 10^{44} \, \rm{erg/s}$ above $z \, {\sim} \, 6$. Our active fraction are significantly larger than the ones reported for JWST sample: which span from $\sim 1 \%$ \citep{Matthee2024} to $\sim 5\% $ \citep{Harikane2023} and $\sim 10\% $ \citep{Maiolino2024}, with higher values for lower luminosity cuts. Note that the active fraction not only strongly depends on the completeness and luminosity cut, but also on the sample considered (effectively, the denominator in the calculation of the fraction). In the calculation of our active fraction, we included all galaxies with an MBH with mass $\mathit{M}_{\rm BH} \, {>} \, 10^{6} \,  \rm{M}_{\odot}$. The active fraction would be lower if we were to include also galaxies with smaller black holes in the sample.

In the bottom panel of Fig. \ref{fig:MBHs_active_fraction} we show again the active fraction, but now considering different mass and $f_{\rm Edd}$ cuts. As expected, we effectively recover the same active fraction of the top panel for black holes with  $\mathit{M}_{\rm BH} \, {>} \, 10^{6} \,  \rm{M}_{\odot}$ and $\mathit{f}_{\rm Edd} \,{>}\, 0.1$ as all these active MBHs shine at luminosities $\mathit{L}_{\rm bol} \, {>} \, 10^{44} \, \rm{erg/s}$. Although, in general, the active fraction decreases with increasing black hole mass, this trend does not hold at the highest redshifts considered in this study. Specifically, at $z \,{\sim}\, 6\,{-}\, 8$ even the most massive black holes show a very large active fraction with an Eddington ratio $\mathit{f}_{\rm Edd} \, {>} \, 0.1$. Interestingly, the \supereddington{} and \heavyMax{} show similar trends.

We now move to the analysis of scaling relations ($\rm L_{bol}\,{-}\, M_{BH}$ and ${ M_{\rm BH}\,{-}\, M_{*}}$) for the two models at $z  \,{=}\, 5$, i.e the typical redshift unveiled by JWST \citep{Maiolino2024,Harikane2023}. These predictions are presented in Fig. \ref{fig:scaling_relations_SE_and_HeavyMax}. The left panel shows the relation between luminosity and black hole mass. Both \supereddington{} and the \heavyMax{} models show that black holes are broadly divided into two populations: one growing close to or above the Eddington limit, and another highly sub-Eddington.  In the \supereddington{} model, MBHs can accrete at rates up to several times the Eddington limit. These super-Eddington episodes occur primarily in moderate-mass black holes ($10^5 \,{\lesssim}\, M_{\rm BH} \,{\lesssim}\, 10^6\,\msun$), as will be further discussed later in the Section. Conversely, for the \heavyMax{} model, these moderate-mass MBHs are mainly accreting at a fraction of the Eddington limit ($0.01 \,{<}\, \mathit{f}_{\rm Edd}\,{<} \, 0.1$). Therefore, when the model predictions are compared with the AGN detected by JWST, we see that the data are compatible with MBHs accreting close to the Eddington limit.

In the middle panel of Fig.~\ref{fig:scaling_relations_SE_and_HeavyMax} we show the $\mathrm{M_{\rm BH}}\,{-}\, \mstar$ relation for active MBHs, defined as the ones shining with a luminosity  $\mathit{L}_{\rm bol} \, {>} \, 10^{44} \, \rm{erg/s}$. We compare again our predictions with JWST results \citep{Maiolino2024,Harikane2023} and  high-$z$ quasars seen by ALMA \citep{Izumi2019}. We see some differences between the two models. For the \supereddington{} case, all MBHs with a luminosity $\mathit{L}_{\rm bol} \, {>} \, 10^{44} \, \rm{erg/s}$ live in galaxies with a narrow range of masses ($10^{8}  \, \rm \msun\, 
 {\lesssim} \, \mstar \, {\lesssim} \, 10^{10.5}  \, \rm \msun\, $), while the range of stellar masses is broader for the  \heavyMax{} model, extending towards lower masses ($10^{7}  \, \rm \msun\, 
 {\lesssim} \, \mstar \, {\lesssim} \, 10^{10.5}  \, \rm \msun\, $). 
Despite these differences, both models produce MBH populations that span a wide range in the $\mathrm{M_{\rm BH}}\,{-}\,M_*$ scaling relation, with the majority lying between the $1\,{:}\,10^2$ and $1\,{:}\,10^3$. However, a non-negligible fraction reaches values as high as a $1 \, {:} \, 1$ ratio, consistent with the high ratios also found for JWST MBHs at those redshifts. The large scatter for both models indicates that active MBHs and their galaxies are still in the process of growth, and a tighter relation is established only at the end of active phases (see also Fig.~\ref{fig:ScalingRelations_All_Models}). When comparing model predictions with observations, we find that JWST-detected AGN are consistent with the MBH populations predicted by both models. However, significant differences between the models emerge at $M_{\*} \,{\lesssim}\, 10^{10} \, \rm \msun$, where the \supereddington{} model predicts more massive black holes. As a result, the detection of active MBHs in these massive high-$z$ galaxies could serve as a key observational discriminator between the two scenarios. Finally, the substantial scatter in the $\mathit{M}_{\rm BH}{-}\mstar$ relation predicted by our models is consistent with recent observational findings. Finally, several studies have proposed that the black hole mass may correlate more fundamentally with the dynamical mass of the host galaxy (rather than with stellar mass alone) as it better captures the galaxy total baryonic content \citep[][]{Maiolino2024, Izumi2019}. Motivated by this, we explore the $\mathit{M}_{\rm BH}\,{-}\,\mathit{M}_{\rm dyn}$ relation in the right panel of Fig.~\ref{fig:scaling_relations_SE_and_HeavyMax}, where we define the dynamical mass as the sum of stellar and gas mass. Although a substantial scatter persists, particularly at the low-mass end, we find evidence for a tighter correlation at higher masses. Notably, only a small number of systems exceed the $1\,{:}\,10^2$ ratio, consistent with observational results from both JWST \citep{Maiolino2024} and ALMA \citep{Izumi2019}.

\begin{figure}
    \centering
    \includegraphics[width=1.0\columnwidth]{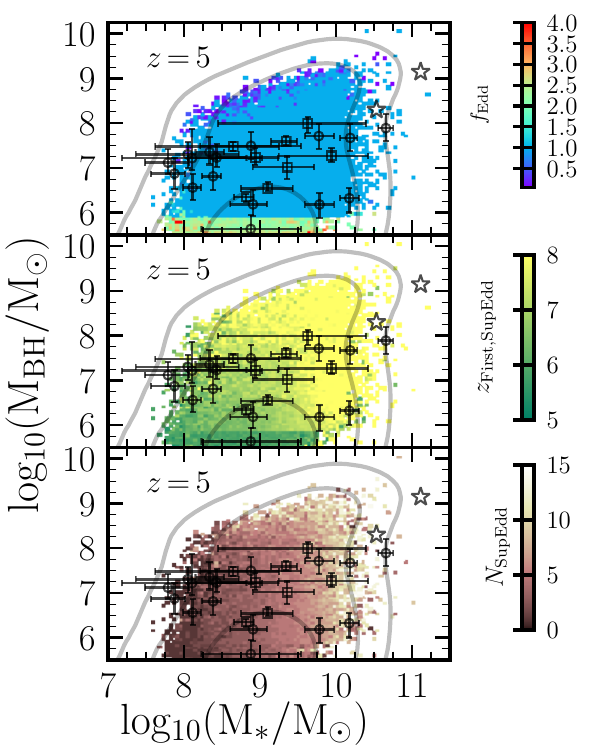}

    \caption{Scaling relation at $z\,{=}\,5$ for MBHs with $\rm L_{bol}\,{>}\,10^{44} erg/s$ for the \supereddington{} model. In each panel, each pixel of the $\rm M_{BH}\,{-}\,M_{Stellar}$ plane is color-coded based on: Eddington ratio of the MBH population computed at $z\,{=}\,5$; redshift of the first super-Eddington event experienced by the $z=5$ MBH population (middle panel); number of super-Eddington phases experienced between formation and until $z\,{=}\,5$ (lower panel). The results are presented for \lgbh{} run on the \MSG{}. The data symbols correspond to the legend presented in the previous plot about the ${M_{\rm BH}\,{-}\,M_{*}}$.}
    \label{fig:scaling_relations_SE}
\end{figure}

To further investigate how super-Eddington accretion influences the location of MBHs in the $\mathit{M}_{\rm BH}\,{-}\,\mstar$ plane, we focus in Fig.~\ref{fig:scaling_relations_SE} on the most luminous MBHs ($\mathit{L}_{\rm bol}  \,{>}\,10^{44}\, \rm erg/s$) predicted by the \supereddington{} model. Instead of showing individual points, here we show the average properties within each pixel on the plane. The top panel is color-coded based on the instantaneous Eddington rate at $z\,{=}\,5$. As shown, more massive black holes have already completed super-Eddington growth and are now growing sub-Eddington. The least massive black holes are instead the ones growing close to Eddington or above. This is consistent with what has been found by  \cite{Maiolino2024} in their JADES sample, where smaller MBHs have generally higher Eddington rates (reaching super-Eddington ones) with respect to more massive ones. In the second panel of Fig.~\ref{fig:scaling_relations_SE}, we present the typical redshift of the first super-Eddington phase. The results show that MBHs with the highest masses and/or in the most massive galaxies are the ones that start growing earlier, probably an indication of a favourable position within the cosmic web for efficient gas cooling, mergers and feeding. The smaller MBHs shown here are just starting to experience super-Eddington growth. Finally, the pixel color-coded of the bottom panel of Fig.~\ref{fig:scaling_relations_SE} represents the average number of Super-Eddington episodes. The model shows that MBHs which started experiencing super-Eddington growth at high redshift have also generally experienced a large number of episodes, even up to ${\sim} \, 10\,{-}\,15$. To a further exploration of the super-Eddington phase, we refer the reader to Appendix~\ref{sec:Activity_And_Duty_Cycle}, where we present the typical duty cycle of the super-Eddington phases. In brief, these are generally short phases, significantly shorter than the phases near or at the Eddington rate. All these results highlight that super-Eddington phases are key in the assembly of the most massive MBHs at high-$z$ if a physically-consistent seeding model is considered \citep[see similar results in][]{trinca2024}.

\subsection{The merging MBHB population}

\begin{figure}
    \centering
    \includegraphics[width=1.0\columnwidth]{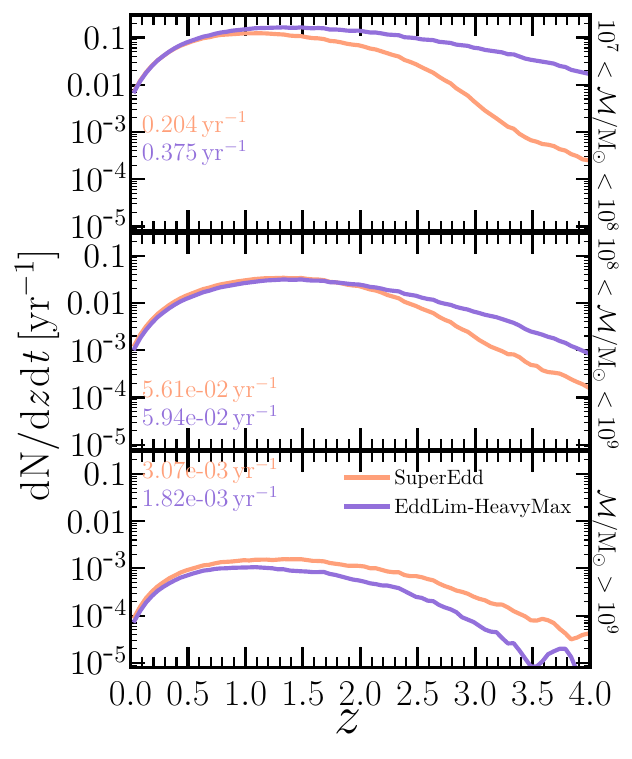}
    \caption{Merger rate of MBHBs with chirp mass $\rm {>}\,10^7\, \msun$ for the \supereddington{} (orange) and \heavyMax{} (purple) models. Each panel corresponds to a given mass bin: $\rm 10^7\,\msun\,{<}\,\mathcal{M}\,{<}\,10^8\, \msun$ (top) $\rm 10^8\,\msun\,{<}\,\mathcal{M}\,{<}\,10^9\, \msun$ (middle) and $\rm \mathcal{M}\,{>}\,10^9\, \msun$ (bottom). The actual total values for the merger rates, for each model and each mass bin, are also quoted in the plot. The numbers quoted in each panel correspond to the total merger rate for each chirp mass bin.}
    \label{fig:MergerRates}
\end{figure}

We now compare the \supereddington{} and \heavyMax{} models in terms of the MBHB populations contributing to the sGWB within the PTA frequency band. Our analysis focuses on binaries with chirp masses $\mathcal{M} \,{>}\, 10^7 \, \rm{M}_{\odot}$, which are expected to dominate the GW signal in this regime. The differences between the two models are characterized by the redshift-dependent merger rate of MBHBs, $dn/dz$, which quantifies the number of mergers per unit redshift and comoving volume. Since \lgbh{} evolves a finite comoving volume, we directly extract this information from the SAM. The quantity $dn/dz$ can be converted into a cosmic merger rate across the observable universe via:
\begin{equation}
\frac{d\mathrm{N}}{dzdt} = \left[\frac{4\pi \, c \, d_L^2}{(1+z)^2}\right] \left(\frac{dn}{dz}\right),
\end{equation}
where $z$ is the merger redshift, $c$ is the speed of light, and $d_L$ is the luminosity distance to the source.\\

The merger rates of our predicted population of MBHBs are shown in Fig.~\ref{fig:MergerRates}. For systems with masses $10^{7}\,{<}\,\mathcal{M}\,{<}\,10^{8}\, \msun$, the \supereddington{} and \heavyMax{} models predict similar merger rates at low-$z$. However, at $z\,{>}\,1$, the \heavyMax{} model indicates a higher number of mergers, with the differences increasing with redshift and reaching up to a factor of hundreds at $z\,{\sim}\,4$. A similar trend is observed for systems with chirp masses in the range $10^{8}\,{<}\,\mathcal{M}\,{<}\,10^{9}\, \msun$, although the differences between the models are less pronounced and begin to emerge only at higher redshifts ($z\,{\sim}\,2$). For systems with $\mathcal{M}\,{>}\,10^{9}, \msun$, the behaviour diverges from that seen in the lower mass bins. Specifically, the \supereddington{} model predicts significantly higher merger rates than the \heavyMax{} model. These differences are mainly concentrated at $z\,{>}\,1$, where the \supereddington{} model yields rates approximately 2–3 times higher. However, unlike in the lower mass regimes, some differences in the model predictions persist down to $z\,{<}\,1$. This behaviour may have important implications for PTAs in the search for continuous GW (CGW) signals, as the first individually resolvable sources are expected to have $\mathcal{M}\,{>}\,10^{9}\, \msun$ and reside at $z\,{<}\,2$ \citep{Rosado2015,Kelley2017,Truant2025}. Consequently, the elevated merger rate predicted by the \supereddington{} model at high chirp masses could result in a larger number of CGW detections compared to the \heavyMax{} scenario.

\begin{figure}
    \centering
    \includegraphics[width=1.0\columnwidth]{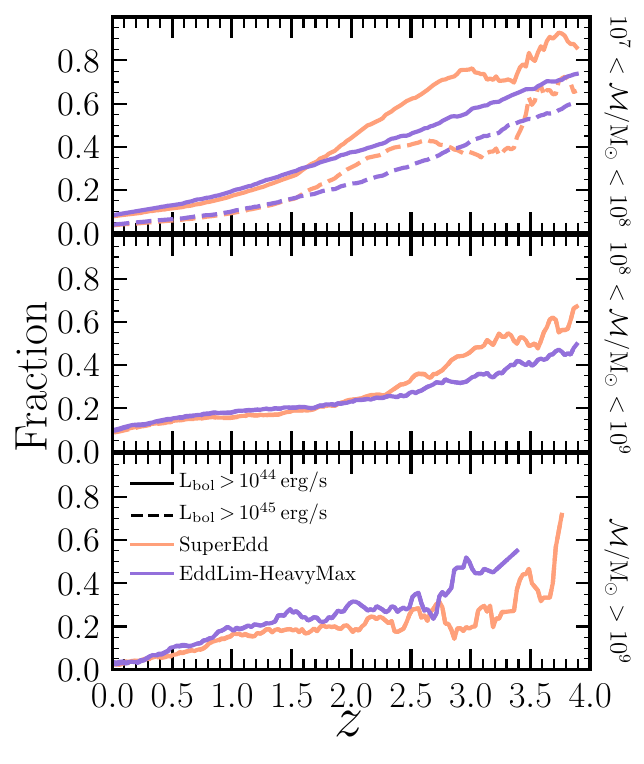}
    \caption{Fraction of merging MBHBs with chirp mass ${>}\,10^7\, \msun$ that satisfy a given luminosity cut (${>}\,10^{44}\, \rm erg/s$, solid and ${>}\,10^{45}\, \rm erg/s$, dashed). Each panel corresponds to a given mass bin: $\rm 10^7\,\msun\,{<}\,\mathcal{M}\,{<}\,10^8\, \msun$ (top) $\rm 10^8\,\msun\,{<}\,\mathcal{M}\,{<}\,10^9\, \msun$ (middle) and $\rm \mathcal{M}\,{>}\,10^9\, \msun$ (bottom).  While orange lines correspond to the \supereddington{} model, the purple ones represent the \heavyMax{} one. We stress that in the middle and bottom panels, the dashed line does not appear because it overlaps with the solid one.}
    \label{fig:Mergers_Luminosity}
\end{figure}

On top of the merger rates, it is interesting to consider the predictions of the two models regarding the potential for conducting multimessenger studies on individual binary systems detected by PTAs \citep[we refer to][for further analysis on this topic]{Truant2025b}. To explore this, in Fig.~\ref{fig:Mergers_Luminosity}, we show the number of MBHB mergers that could be associated with an electromagnetic (EM) counterpart with a bolometric luminosity brighter than $\rm 10^{44}/10^{45}\, erg/s$. For binaries with masses of $10^{7}\,{<}\,\mathcal{M}\,{<}\,10^{8}\, \msun$, both the \supereddington{} and \heavyMax{} models show an increasing trend towards higher redshifts. Despite these similarities, some differences between the models are evident. At $z\,{\sim}\,3$, \supereddington{} predicts that 80\% of these mergers will have an EM emission, whereas \heavyMax{} estimates about 60\%. These differences diminish at lower redshifts ($z\,{\leq}\,2$), where both models forecast fractions below 30\%. Similar trends are observed for merging systems with masses of $10^{8}\,{<}\,\mathcal{M}\,{<}\,10^{9}\, \msun$. Finally, the most significant differences are found in systems with $\mathcal{M}\,{>}\,10^{9}\, \msun$. Specifically, the \heavyMax{} model predicts a higher fraction of systems with an electromagnetic (EM) counterpart of $\rm \mathit{L}_{\rm bol}\,{>}\,10^{44}, erg/s$ at $z\,{>}\,1.5$. For example, at $z\,{\sim}\,2$, \heavyMax{} estimates that approximately 30\% of merging binaries could have detectable EM counterparts, compared to about 20\% in the \supereddington{} model. These modest differences persist down to $z\,{\sim}\,0.5$, but disappear by $z\,{\sim}\,0.25$, where both models predict fewer than 5\% of systems with such luminous EM emission.\\

The results presented above highlight that different black hole seeding mechanisms and growth models can produce significant variations in both the merger rates and EM counterparts of MBHBs within the PTA frequency band. In particular, our analysis shows that differences in the population of high-mass systems ($\mathcal{M}\,{>}\,10^{9}\, \msun$) can persist down to $z\,{<}\,0.5$. This is especially important, as massive, low-redshift binaries represent prime targets for CGW searches with PTAs \citep{Rosado2015,Truant2025b}. Therefore, our findings suggest that in the coming years, as PTAs reach the sensitivity required for CGW detection, it will become feasible to constrain models of MBH formation and evolution through PTA observations. Moreover, the larger differences observed at high redshift for MBHBs with moderate masses make future GW detectors such as LISA particularly promising for distinguishing between these models, especially when combined with electromagnetic observations in a multimessenger framework. Finally, the latest JWST observations have begun to constrain the fraction of dual AGNs at high redshifts \citep[][]{Maiolino2024,Perna2025}, which likely represent the progenitors of future MBHBs. As the sample of high-$z$ dual AGNs continues to grow, it will become increasingly feasible to place meaningful constraints on MBH seeding and growth models, particularly since our results suggest that the most significant differences between models arise at high redshift. In future work, we will investigate how the visibility and abundance of dual AGNs are influenced by different assumptions about MBH formation and accretion.

\section{Summary and conclusions} \label{sec:summary}

In this work, we explore how recent JWST results on the $z\,{>}\,4$ moderate-luminosity AGN population and the sGW background currently observed with PTA experiments, can serve as \textit{statistical multimessenger} probes to constrain the formation and growth of MBHs. Our analysis is based on the theoretical predictions of the \lgbh{} model, an extension of the widely used and tested \lgal{} semi-analytical model for galaxy formation and evolution. The \lgbh{} model, presented in detail here, has been developed through a series of studies to create a comprehensive description of the key physical processes governing MBH evolution, including seeding, spin evolution, growth through multiple channels, dynamical interactions, and mergers \citep[][]{IzquierdoVillalba2020, IzquierdoVillalba2021, Spinoso2022, Polkas2024}. In this paper we also introduce a novel methodology that combines the outputs of the \lgbh{} model when applied to the merger trees of both the \texttt{Millennium} and \texttt{Millennium-II} simulations, leveraging the large cosmological volume of the former and the higher resolution of the latter. The combination of the outputs from both simulations is done through our \graft{} procedure and allows us to derive predictions for both moderate mass as well as the most massive black holes from their early formation stages down to the local Universe within the same cosmological volume.\\

We apply this framework to investigate four  variations of MBH formation and growth, focusing on their predictions for the high-redshift AGN population and the gravitational wave background from MBH binary mergers. Specifically, we compare (i) the \supereddington{} model, which features a physically-motivated MBH seeding, including light and heavy seeds, and allows for phases of super-Eddington accretion when conditions are favorable, (ii) the \eddington{} variation, which follows the same seeding procedure of the \supereddington, but growth is capped at the Eddington rate, the
(iii) the \heavyMin{} and (iv) the \heavyMax{} models, in which  all seeds are artificially assumed to be heavy and MBH growth is capped at the Eddington limit.  The \heavyMax{} adopts a very high heavy seed occupation fraction and \heavyMin{} a lower one, but they both exceed the number densities predicted even by the most optimistic physical scenarios for direct-collapse black holes. 
Our analysis shows that all four models predict a low-redshift ($z\,{<}\,2$) population of black holes and quasars that is consistent with key observational constraints, including the local black hole mass function, the AGN luminosity function, and the $\mathit{M}_{\rm BH}\,{-}\,\mstar$ scaling relation. At these redshifts, some differences between the models emerge in the low-mass regime ($M_{\rm BH} \,{<}\,10^6\, \msun$, $L_{\rm bol} \,{<}\,10^{44}\, \rm erg/s$) but current observational data are not yet sufficient to constrain this part of the parameter space. We thus turn to the recent high-redshift AGN observations from JWST and the sGWB measurements from PTA experiments to break the degeneracies between these models. On the JWST side, we are working under the assumptions that the recent AGN observations are not significantly affected by observational biases and systematics, and that all LRDs are accreting MBHs (i.e., we are not making any color selection in our analysis and we refer to Herrero-Carrión, in prep., for an analysis of the LRD colors predicted by \lgbh{}). On the PTA side, we assume that the entire PTA signal originates from an astrophysical population of merging circular MBH binaries. Given these assumptions, we can draw the following conclusions:
\begin{enumerate}
    \item[-] \underline{Constraints from JWST:}  When comparing the predictions of our models with recent estimates of the AGN  \textit{bolometric luminosity function} and the \textit{MBH mass function} at $z\,{>}\,4$, derived from JWST surveys \citep[][]{Matthee2024, Akins2024, Kokorev2024, Maiolino2024, Greene2024, Geris2025, Greene2025}, we find that the \eddington{} model underpredicts the observed number densities of AGN. Indeed, JWST observations indicate the existence of a large population of moderately luminous, high-redshift AGN, that cannot be accounted for without invoking either episodes of super-Eddington accretion (\supereddington{} model) or an extremely high number density of heavy seeds (\heavyMax{} and \heavyMin{} models). Such mechanisms appear necessary to assemble a significant population of massive black holes by $z \,{\sim} \,5$.\\
    
    \item[-] \underline{Constraints from PTA:} We computed the expected sGWB amplitude at the reference frequency of $1 \, \mathit{yr}^{-1}$ for all four models. All, except the \heavyMin{} model, predict amplitudes that are consistent with the current observational constraints reported by the EPTA, NANOGrav, PPTA, and CPTA collaborations \citep[][]{Antoniadis2023, Agazie2023, Reardon2023, Xu2023}. These PTA measurements favor models with a high occupation fraction of MBHs and a number density of MBHs with $\mathit{M}{\rm BH} \,{\gtrsim}\,10^{9} \, \rm M{\odot}$ that is consistent with (or slightly exceeds) the most recent estimates of the high-mass end of the black hole mass function \citep[e.g.,][]{LiepoldAndMa2024}, as also noted by \citet{IzquierdoVillalba2021}. Furthermore, the clear tension between the \heavyMin{} predictions and current PTA data suggests that models with moderate heavy seed formation efficiency are disfavored. 
\end{enumerate}

The combined constraints from JWST and PTA observations indicate that the only models capable of simultaneously reproducing both datasets are the \supereddington{} and \heavyMax{} scenarios. These two models are uniquely able to produce a sufficiently large population of MBHs already in place by $z \,{\sim}\, 5\,{-}\,7$, as required by JWST observations, while also predicting a high number density of very MBHs at lower redshifts, consistent with PTA constraints on the sGWB. This result is broadly in agreement with findings from other studies based on semi-analytical models and parametrized approaches \citep[e.g.,][]{Toubiana2024}, reinforcing the conclusion that enhanced MBH seed formation and accelerated growth are necessary to reproduce the observed MBH population across cosmic time. We further examine additional observable quantities predicted by these two preferred models. In particular, we focus on the single and binary MBH population:

\begin{itemize}
    \item \textbf{Single MBHs}: Both models predict a nearly constant \textit{black hole accretion rate density} between $z \,{\sim}\, 2$ and $5$, although the \heavyMax{} model exhibits a peak at slightly higher redshift compared to the \supereddington{} model. The \textit{black hole mass densities} are very similar in both models and consistent with recent observational estimates. However, the \supereddington{} model predicts a larger mass assembly at high redshifts ($z \,{\sim}\, 7\, {-} \,8$). This difference is also reflected in the \textit{active fraction of black holes}, where the \supereddington{} model forecasts a generally higher fraction of active MBHs across all redshifts, particularly pronounced at $z \,{>}\,6$. Correspondingly, it predicts a greater number of very luminous AGN ($\rm \mathit{L}_{bol} \,{\gtrsim}\, 10^{46} \, erg\,s^{-1}$) than the \heavyMax{} model at $z \,{>}\, 5$. Conversely, the heavy seed only models tend to produce a larger number density of moderate luminosity AGN ($\rm \mathit{L}_{bol} \,{\lesssim}\, 10^{45} \, erg,s^{-1}$). Indeed, the two models predict a different shape of the luminosity function, with the heavy-boosted model predicting a steeper slope. If the number density of AGN at the highest luminosities $\rm \mathit{L}_{bol} \,{\gtrsim}\, 10^{46} \, erg,s^{-1}$ is confirmed as high as $\sim 10^{-5} \, \rm{Mpc}^{-3}$ \citep[but see][]{Greene2025}, other ``exotic'' channels of MBH formation would be need to be invoked (e.g., primordial black holes). 
Finally, when looking at the \textit{ ${\rm M_{BH}}\,{-}\,M_{*}$ scaling relation}, both models exhibit a large scatter, especially for luminous and less massive black holes which are still in the process of intense growth. Consistent with JWST observations, both predict a significant population of MBHs residing above the local relation, as well as a notable fraction with masses comparable to those of their host galaxies.\\
    
    \item \textbf{Binary MBHs}: Significant differences arise between the \supereddington{} and \heavyMax{} models when comparing the \textit{merger rates} of MBHBs with chirp masses $\mathcal{M} \,{>}\, 10^7 \, \msun$ at $z \,{>}\, 2$. These differences diminish at lower redshifts for systems with $\mathcal{M}  \,{<}\, 10^9 \, \msun$. However, the \supereddington{} model predicts notably higher merger rates for $\mathcal{M} \,{>}\, 10^9 \, \msun$ MBHBs in the redshift range $1 \,{<}\, z \,{<}\, 2$. Such differences in the population of the most massive MBHBs have important implications for PTAs in their search for continuous gravitational wave (CGW) signals, suggesting that future CGW detections with PTAs will provide valuable constraints on MBH seeding and growth models. As we will explore in a dedicated paper, we can also foresee that LISA will be able to play a crucial role in discriminating models of seeding and growth of MBH, being particularly sensitive in the $10^5\,{<}\,<\mathcal{M} \,{<}\, 10^7 \, \msun$ chirp mass range.
\end{itemize}

Although the current observational constraints employed in this work already place meaningful limits on MBH formation models, they are not yet sufficient to fully discriminate between different seeding and growth scenarios. We expect that this will happen when the determination of the AGN luminosity function at $z \,{\gtrsim}\, \, 5$ will become more precise. Additionally, studies of the environments and spatial distribution of AGNs offer valuable avenues for model differentiation. In a forthcoming work, we plan to investigate the properties of host galaxies and the large-scale environments of MBHs within the \supereddington{} and \heavyMax{} models to provide further constraints. However, it is important to emphasize that the \heavyMax{} model assumes an artificially elevated number density of heavy seeds, significantly exceeding even the most optimistic theoretical predictions to date.  \\

To conclude, the \textit{statistical} multi-messenger analysis presented in this work underscores the necessity of adopting a holistic approach to modeling the co-evolution of galaxies and MBHs. Robust constraints can only be achieved by integrating diverse observational data across cosmic time with physically motivated models embedded within a full cosmological framework. In this context, \lgbh{} emerges as a state-of-the-art semi-analytical model that uniquely combines detailed models for galaxy and black hole formation and evolution. Furthermore, through its \graft{} procedure, \lgbh{} uniquely combines self-consistently the strengths of both high-resolution/small-volume and low-resolution/large-volume simulations, enabling predictions over a wide dynamical range, from high to low redshifts, and across both electromagnetic and gravitational wave observational windows.

\begin{acknowledgements}

S.B. acknowledges support from the Spanish Ministerio de Ciencia e Innovaci\'on through project PID2021-124243NB-C21 and the Alexander von Humbold Foundation via a Research Fellowship for support during research stays at the Max Planck Institute for Astrophysics. D.I.V and A.S acknowledge the financial support provided under the European Union’s H2020 ERC Consolidator Grant ``Binary Massive Black Hole Astrophysics'' (B Massive, Grant Agreement: 818691) and the European Union Advanced Grant ``PINGU'' (Grant Agreement: 101142079). D.S. acknowledges support by the Fondazione ICSC, Spoke 3 Astrophysics and Cosmos Observations. National Recovery and Resilience Plan (Piano Nazionale di Ripresa e Resilienza, PNRR) Project ID CN\_00000013 "Italian Research Center on High-Performance Computing, Big Data and Quantum Computing" funded by MUR Missione 4 Componente 2 Investimento 1.4: Potenziamento strutture di ricerca e creazione di "campioni nazionali di R\&S (M4C2-19 )" - Next Generation EU (NGEU). M.C. acknowledges funding from MIUR under the grant PRIN 2017-MB8AEZ, from the INFN TEONGRAV initiative, and from the MUR Grant "Progetto Dipartimenti di Eccellenza 2023-2027” (BiCoQ). D.S. acknowledges support by the Fondazione ICSC, Spoke 3 Astrophys
   
\end{acknowledgements}

\bibliographystyle{aa.bst} % style aa.bst
\bibliography{references.bib} % your references Yourfile.bib

\appendix

\section{Activity and duty cycle} \label{sec:Activity_And_Duty_Cycle}

\begin{figure}
    \centering
    \includegraphics[width=1.0\columnwidth]{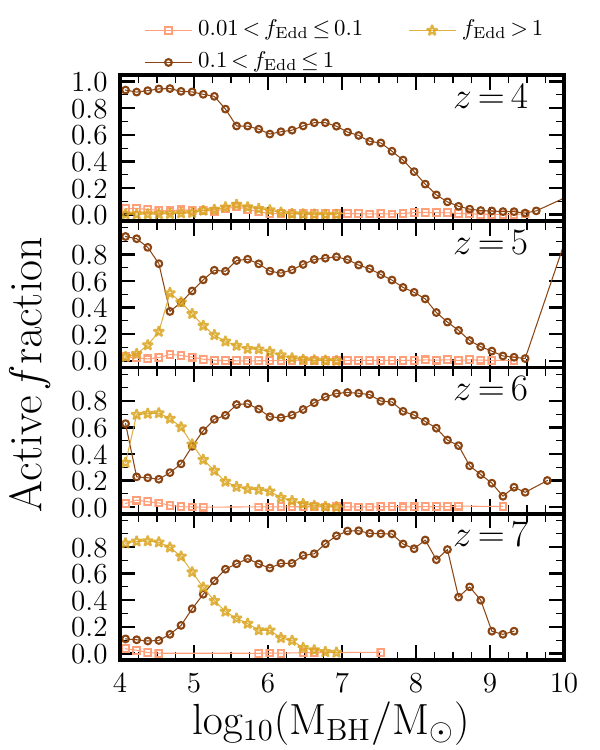}
    \caption{Fraction of active MBHs over the entire population, split into three accretion regimes according to their Eddington-ratio ($f_{\rm Edd}$). Results are shown for the \supereddington{} model, at  $z\,{=}\,4$, $z\,{=}\,5$, $z\,{=}\,6$ and $z\,{=}\,7$ (respectively from top to bottom).}
    \label{fig:ActiveFraction}
\end{figure}

In Fig.~\ref{fig:ActiveFraction} we present the fraction of MBHs accreting in a certain range of $f_{\rm Edd}$ values. At $z\,{=}\,7$ and $z\,{=}\,6$, the model predicts that ${\sim}\,70\%$ of the MBH population with with $\rm 10^{4}\,{<}\,M_{BH}\,{<}\,10^5 \, \msun$ is experiencing a super-Eddington accretion, with values that range between $1\,{<}\,f_{\rm Edd}\,{\leq}\,3$. At $\rm 10^{5}\,{<}\,M_{BH}\,{<}\,10^7 \, \msun$, the fraction of MBHs accreting in this mode decreases down to ${\leq}\,10\%$ while those growing at Eddington ratios of $0.01\,{<}\,f_{\rm Edd}\,{\leq}\,1$ account for approximately 80\%. For more massive systems, the Super-Eddington growth model is no longer present, and the contribution of MBHs with $0.01\,{<}\,f_{\rm Edd}\,{\leq}\,1$ decreases as the mass increases, favoring a non-accreting mode. At $z\,{=}\,5$ the behavior at masses ${>}\,10^7 \msun$ remains similar. However at  $\rm 10^{4}\,{<}\,M_{BH}\,{<}\,10^5 \, \msun$, important differences are seen. Specifically, the super-Eddington accretion with $1\,{<}\,f_{\rm Edd}\,{\leq}\,3$ represents less than 50\% of the accreting population, being the accretion mode at $0.01\,{<}\,f_{\rm Edd}\,{\leq}\,1$ the one which dominates. Finally, at $z\,{=}\,4$ the fraction of population with $f_{\rm Edd}\,{>}\,1$ decreases, especially at high masses. Furthermore, these super-Eddington cases became extremely rare, representing less than 5\% of the population of MBHs with $\rm M_{BH}\,{<}\,10^5 \, \msun$.

\begin{figure}
    \centering
    \includegraphics[width=1.0\columnwidth]{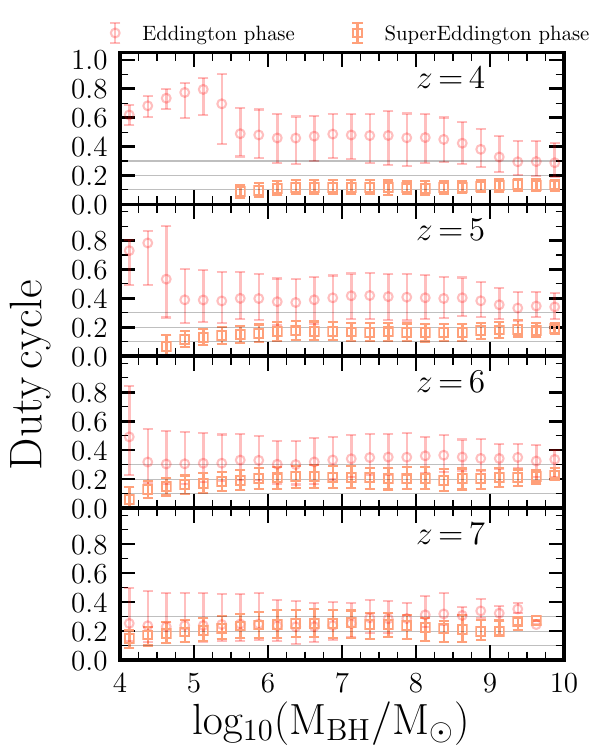}
    \caption{Duty cycle of MBHs computed separately for the Eddington and Super-Eddington, as a function of MBH mass. From top to bottom, we show the distribution of duty-cycles computed for $z\,{=}\,4$, $z\,{=}\,5$ and $z\,{=}\,6$ and $z\,{=}\,7$. Note that we define the duty cycle as the ratio between the total time (i.e. integrated over its entire life-time) spent by an MBH in an Eddington-limited or Super-Eddington active phase versus the total life-time of the MBH.}
    \label{fig:Duty_Cycle_SE}
\end{figure}

As shown in Fig.~\ref{fig:ActiveFraction}, MBHs at $4\,{<}\,z\,{<}\,6$ primarily accrete at the Eddington limit. Super-Eddington accretion events are limited to MBHs with $\rm 10^{4}\,{<}\,M_{BH}\,{<}\,10^5 \, \msun$. Despite that, some MBHs with masses greater than $\rm M_{BH}\,{>}\,10^5 \, \msun$ may have needed to experience super-Eddington episodes to build up a significant portion of their mass. To explore further on this, in Fig.~\ref{fig:Duty_Cycle_SE} we explore the duty cycle of MBHs divided between super-Eddington and Eddington phases. Note that we define the duty cycle as the ratio between the total time (i.e. integrated over its entire life-time) spent by an MBH in an Eddington-limited or Super-Eddington active phase versus the total MBH life-time. As shown, the duty cycle associated with a super-Eddington phase does not vary much with the black hole mass but features a strong redshift dependence. For instance, the super-Eddington duty cycle of objects at $z\,{=}\,7$ and $z\,{=}\,6$ can reach 30\% while at $z\,{=}\,5$ and $z\,{=}\,4$ this values drop down to 20\% and 10\% respectively. The duty cycle associated with Eddington growth shows both a redshift and mass dependence. Overall, the values tend to be higher towards lower redshifts. For instance, at $z\,{=}\,6$ we find duty cycles of 40\% while at $z\,{=}\,4$ the values can reach 60\%. Regarding the mass dependence, the results show that the lower the MBH mass, the larger the duty cycles.\\


%\bibliographystyle{aa.bst} % style aa.bst
%\bibliography{references.bib} % your references Yourfile.bib

\end{document}